\documentclass[sigconf]{acmart}

\AtBeginDocument{%
  \providecommand\BibTeX{{%
    \normalfont B\kern-0.5em{\scshape i\kern-0.25em b}\kern-0.8em\TeX}}}

\usepackage{natbib}
\usepackage{amsmath,amsfonts}
\usepackage{algorithmic}
\usepackage{graphicx}
\usepackage{textcomp}
\usepackage{footmisc}
\usepackage{xcolor}
\usepackage{multirow}
\usepackage{amsmath}
\usepackage{balance}       
\usepackage{graphics}      
\usepackage[T1]{fontenc}   
\usepackage{txfonts}
\usepackage{mathptmx}
\usepackage{booktabs}
\usepackage{textcomp}
\PassOptionsToPackage{hyphens}{url}
\usepackage{microtype}        
\usepackage{ccicons}          

\usepackage{amsmath,amssymb,amsfonts}
\usepackage{algorithmic}
\usepackage{graphicx}
\usepackage{textcomp}
\usepackage{amsmath,amssymb,amsfonts}
\usepackage{algorithmic}
\usepackage{graphicx}
\usepackage{textcomp}
\usepackage[ampersand]{easylist}
\usepackage{amssymb}
\usepackage{multicol}
\usepackage{multirow}
\usepackage{tabularx}
\usepackage{rotating}
\usepackage[roman]{parnotes}
\usepackage[many]{tcolorbox}
\usepackage{soul}

\usepackage{comment}
\usepackage{booktabs}
\usepackage{colortbl}
\usepackage{verbatim}
\usepackage{array}
\usepackage[inline]{enumitem}
\usepackage[export]{adjustbox}
\usepackage{pifont}
\usepackage{mathtools} 
\usepackage{amsmath}
\usepackage{hyphenat}
\DeclareTextFontCommand{\mytexttt}{\ttfamily\hyphenchar\font=45\relax}

\newcommand{\review}[1]{{\color{black}{#1}}}

\setlist[itemize]{noitemsep, topsep=0pt}
\newcommand{\MyBox}[1]{\vspace{3mm}\noindent\framebox[\columnwidth][c]{\parbox[b]{0.95\columnwidth}{ #1 }}\vspace{3mm}}

\newif\ifdraft
\draftfalse 

\makeatother

\def\tablename{Table}
\def\BibTeX{{\rm B\kern-.05em{\sc i\kern-.025em b}\kern-.08em
    T\kern-.1667em\lower.7ex\hbox{E}\kern-.125emX}}

\newcolumntype{L}[1]{>{\hsize=.8\hsize\raggedright\arraybackslash}m{#1}}
\newcolumntype{R}[1]{>{\hsize=.8\hsize\raggedleft\arraybackslash}m{#1}}
\newcolumntype{C}[1]{>{\hsize=.8\hsize\centering\arraybackslash}m{#1}}

\copyrightyear{2022}
\acmYear{2022}
\setcopyright{acmcopyright}\acmConference[ICSE-SEIS'22]{Software Engineering in Society}{May 21--29, 2022}{Pittsburgh, PA, USA}
\acmBooktitle{Software Engineering in Society (ICSE-SEIS'22), May 21--29, 2022, Pittsburgh, PA, USA}
\acmPrice{15.00}
\acmDOI{10.1145/3510458.3513008}
\acmISBN{978-1-4503-9227-3/22/05}

\begin{document}


\title{Perceptions of the State of D\&I and D\&I Initiative in the ASF}
\author{Mariam Guizani}
\affiliation{%
  \institution{Oregon State University}
  \city{Corvallis}
  \state{Oregon}
  \country{USA}}
\email{guizanim@oregonstate.edu}

\author{Bianca Trinkenreich}
\affiliation{%
 \institution{Northern Arizona University}
 \streetaddress{1 Th{\o}rv{\"a}ld Circle}
 \city{Flagstaff, AZ}
 \country{USA}}
\email{bt473@nau.edu}

\author{Aileen Abril Castro-Guzman}
\affiliation{%
 \institution{Oregon State University}
 \city{Corvalis, OR}
 \country{USA}}
\email{castroga@oregonstate.edu}

\author{Igor Steinmacher}
\affiliation{%
 \institution{Northern Arizona University}
 \city{Flagstaff, AZ}
 \country{USA}}
\email{igor.steinmacher@nau.edu}

\author{Marco Gerosa}
\affiliation{%
 \institution{Northern Arizona University}
 \city{Flagstaff, AZ}
 \country{USA}}
\email{marco.gerosa@nau.edu}

\author{Anita Sarma}
\affiliation{%
 \institution{Oregon State University}
 \city{Corvalis, OR}
 \country{USA}}
\email{anita.sarma@oregonstate.edu}

\renewcommand{\shortauthors}{Guizani et al.}

\begin{abstract}

Open Source Software (OSS) Foundations and projects are investing in creating Diversity and Inclusion (D\&I) initiatives. However, little is known about contributors' perceptions about the usefulness and success of such initiatives. We aim to close this gap by investigating how contributors perceive the state of D\&I in their community. In collaboration with the Apache Software Foundation (ASF), we surveyed 600+ OSS contributors and conducted 11 follow-up interviews. We used mixed methods to analyze our data--quantitative analysis of Likert-scale questions and qualitative analysis of open-ended survey question and the interviews to understand contributors' perceptions and critiques of the D\&I initiative and how to improve it. Our results indicate that the ASF contributors felt that the state of D\&I was still lacking, especially regarding gender, seniority, and English proficiency. Regarding the D\&I initiative, some participants felt that the effort was unnecessary, while others agreed with the effort but critiqued its implementation. These findings show that D\&I initiatives in OSS communities are a good start, but there is room for improvements. Our results can inspire the creation of new and the refinement of current initiatives.
\\
\\
\textbf{Lay Abstract:} 
Open Source Software (OSS) is widely used in society (e.g., Linux, Chrome, and Firefox), and contributing to these projects helps individuals learn and showcase their skills, so much so that the history of contributions are increasingly being analyzed by hirers. However, the people who contribute to OSS are predominately men (about 90\%). This means that women and other minorities lose out on job opportunities and OSS projects lose out on diversity of thought. OSS organizations such as the Apache Software Foundation (ASF) promote a variety of initiatives to increase diversity and inclusion (D\&I) in their projects, but they are piece-meal and little is known about contributors' perceptions about the usefulness and success of these initiatives. Here, we surveyed and interviewed ASF contributors to understand their perceptions about the state of D\&I in the ASF and the effectiveness of existing D\&I initiatives. Our findings show that individuals who are in the minority face challenges (e.g., stereotyping, lack of peer-network, and representation in decision making) and \review{contributors' perceptions of the D\&I initiative are a mixed bag, ranging from commending the current efforts to considering them to be ``lip service''.} 
\review{These findings suggest that current D\&I initiatives in OSS communities are a good start, but much needs be done in terms of creating new successful initiatives and refining current ones.}
\end{abstract}
\keywords{Diversity, D\&I initiative, Open Source Software}
\maketitle
\section{Introduction}
Open Source Software (OSS) now plays a key role in software development as well as in workforce development, where contributors join projects to learn new skills~\cite{gerosa2021motivation}, showcase their technical skills~\cite{Marlow2013CSCW}, and improve their career paths ~\cite{Singer2013CSCW, trinkenreich2020hidden}. However, despite its ubiquity, it is well documented that OSS has low diversity~\cite{terrell2017gender,lee2019floss}. 

Low diversity in OSS has unfortunate effects: (1) OSS projects miss out on the benefits of a diverse set of individuals and qualifications, or from the diversity of thought that these potential contributors could bring; (2) developers in the minority miss out on the learning and experience opportunities that OSS projects provide; and (3) job opportunities evade these individuals when OSS contributions are used to make hiring decisions~\cite{Singer2013CSCW, Marlow2013CSCW}. 

Gender diversity in OSS has been widely studied showing women are severely underrepresented. From an analysis of ten OSS projects, \citet{bosu2019diversity} found less than 10\% of contributors to be women and that women were rarely included in leadership positions.
Other studies have reported similar low numbers (e.g.,~\cite{David2008IEP, Ghosh2002IIIM, bosu2019diversity, vasilescu2015gender, robles2016women}). 
Researchers have also investigated how gender can differentiate team dynamics~\cite{vasilescu2015perceptions} and team perceptions \cite{ICSE-Confidence-Competence-2018, vasilescu2015perceptions, Lee.Carver:2019}, career progression~\cite{canedo2020work}, motivations to join~\cite{gerosa2021motivation} and type of contributions ~\cite{trinkenreich2020hidden}. Others have reported on barriers that women face because of the technologies in use \cite{mendez2018open}, in getting their contributions accepted \cite{Terell-2017} and participating in a generally hostile environment ~\cite {nafus2012patches,  prana2020including}.

More recently, research has started to look at other diversity attributes. For example, studies have investigated the effects of location and language on getting contributions accepted ~\cite{furtado2020successful, rastogi2016geographical, rastogi2018relationship}. Others have analyzed age diversity by analyzing its impact on code reviews \cite{murakami2017wap}, or motivations to continue to contribute~\cite{davidson2014older, morrison2016veteran}. The majority of research has focused on only one attribute of diversity, with a few emerging works extending the gender diversity analysis by incorporating another diversity attribute (e.g., location \cite{ortu2017diverse, prana2020including}, cognitive style~\cite{padala2020gender}). 
Diversity, however, is a multi-dimensional construct that arises from attributes that differentiate people, demographic (e.g., age, gender, ethnicity) or otherwise (e.g., role, expertise, personality, cognitive styles) and focusing on a single attribute as the analysis lens, gives only a partial view of this complex phenomenon. Therefore, in this work we investigate:

\textit{(RQ1.) How do contributors with different backgrounds perceive the state of D\&I in their OSS community?}\\
The goal of RQ1 is to build and extend the current literature by analyzing the following six contributors background attributes: gender, education, English proficiency, seniority at ASF, compensation type and country of residence.
We build on work by \citet{Lee.Carver:2019} who investigated contributor perception of the state of D\&I using the gender lens. Our study explores participant perceptions of role stereotyping and their ability to contribute using questions from ~\cite{Lee.Carver:2019}. 
While gender, country of residence and English proficiency attributes have been investigated by others, either in isolation or along with another lens, we are unaware of literature investigating education and compensation type. Given the changing landscape of OSS \cite{Steinmacher.Teenager:2017} where companies are increasingly employing developers to contribute to OSS and people are seeing OSS as a career stepping stone \cite{trinkenreich2020hidden, gerosa2021motivation}, its important to understand how these attributes play a role in creating inclusive, diverse OSS communities. 

OSS projects are well aware of problems of toxic interactions that create a non-inclusive environment and low gender diversity~\cite{Lee.Carver:2019}. To overcome these issues OSS projects often include a code of conduct~\cite{singh2019open} to manage communication expectations. Several OSS foundations have also started broader D\&I initiatives
such as, the Linux foundations' Software Developer Diversity and Inclusion (SDDI) project ~\cite{linuxFoundationDAndI},  the D\&I working group of the Community Health Analytics Open Source Software (CHAOSS) project ~\cite{chaossmetrics}, and the D\&I initiative at the ASF~\cite{EDIgroup}. These initiatives are tasked to improve diversity in different aspects of OSS such as, diversity in events, inclusive naming, and code of conduct. 

While these OSS initiatives are important in their goals of fixing specific issues (e.g., toxic interactions, non-inclusive naming), they are siloed and we lack an understanding of how the community perceives them. Such an understanding is important to realise what is working and what is not to guide the initiatives.  

\textit{(RQ2.)  How do contributors perceive D\&I initiative(s) in their OSS communities?}\\
We answer these research questions by partnering with the ASF Diversity and Inclusion (D\&I) committee~\cite{EDIgroup} because of their interest in understanding their contributors' perspective of the D\&I state and the D\&I initiative. We conducted an online survey with 600+ ASF respondents. The survey was designed in collaboration with the D\&I committee and the ASF community at large. We followed up the survey results with 11 interviews to get a deeper understanding of contributors' perspectives. We used mixed methods to analyze the survey and interview responses. Our results shed light on contributors' background attributes that influence their perception of the state of D\&I (see section \ref{sec:RQ1}) and their perception of the ASF D\&I initiative, from its necessity to its efficacy and where it can be improved (see section \ref{sec:RQ2}).

\vspace{-2pt}
\section{Background and Related Work}
Diversity in OSS has gained considerable attention in recent years with OSS projects and foundations investing in efforts to create diverse and more inclusive communities. 
Research has also investigated the topic of low diversity and barriers to contributing to OSS. A majority of which has focused on one diversity aspect--gender, investigating gender distribution of women in OSS \cite{bosu2019diversity, robles2016women, ortu2017diverse, lin2016recognizing, gila2014impact, izquierdo2018openstack, robles2014floss} and in leadership positions~\cite{canedo2020work}, perceptions of women contributors in OSS \cite{Lee.Carver:2019, vasilescu2015perceptions}, the impact of gender on productivity \cite{vasilescu2015gender} and the barriers that women face \cite{Terell-2017,  nafus2012patches, mendez2018open, prana2020including, ICSE-Confidence-Competence-2018}.

For instance, \citet{vasilescu2015perceptions} used the gender lens to understand GitHub contributors' perception of their team and awareness of their teammates' backgrounds, with gender being the second-most noticed attribute.  Other research focused on women's experience in OSS and support systems in place to increase women's participation \cite{singh2019women, singh2019open}. 
\citet{singh2019open} found that only 12 out of 355 OSS websites have `women only' sections and \citet{Lee.Carver:2019} found that, while some contributors expressed a positive feeling about women’s participation in OSS, some were strongly opposed to their inclusion. 
Finally, researchers have investigated barriers that women face in tools and technology~\cite{mendez2018open, padala2020gender}, in getting pull requests accepted~\cite{Terell-2017}, and participating in discussions ~\cite{nafus2012patches, prana2020including}. 

Research has investigated the experience of ``older'' contributors in OSS \cite{murakami2017wap, morrison2016veteran, davidson2014older}. For instance, \citet{murakami2017wap} looked at how age can impact code reviews and found that age has no significant effect on code review correctness and efficiency. \citet{morrison2016veteran} investigated the low participation of veteran software developers in OSS and how their contributions differ from those of their younger peers. \citet{morrison2016veteran} results reflected that veteran OSS contributors are less socially motivated than their younger counterparts, which aligns with \citet{davidson2014older} findings that older contributors face more social than technical challenges. 

The impact of location on pull request acceptance has also been investigated. \citet{furtado2020successful} found that contributors from countries with low human development indexes face the most pull request rejections. Similarly, \citet{rastogi2018relationship} investigated the top countries with highest and lowest pull request acceptance rates and \citet{rastogi2016geographical} found pull request acceptance rate increases by 19\% when the submitter and integrator are from the same country. 

Recent studies have started to investigate diversity through the lens of multiple demographic attributes. For example, \citet{prana2020including} investigated the difference in gender diversity between geographic regions and found that there has been a small improvement of gender diversity amongst contributors in Northern America and South-Eastern Asia.
\review{\citet{catolino2019gender} investigated community smells through the gender and experience lenses.}
\citet{ortu2017diverse} also used a dual-lens approach and found that gender diversity increased productivity, while intra-team nationality diversity decreased the level of politeness.

In summary, existing research has largely investigated the topic of diversity using a single lens (e.g., gender, location, or age). Recently, emerging studies have started investigating diversity by combining gender with a second lens. Our work complements these works by taking a multi-dimensional approach to investigate contributors' perception of the state of diversity using 6 demographic lenses---\review{gender, education, English proficiency, seniority at ASF, compensation type and country of residence---that can impact contributors' experiences in their OSS community.}
We also investigate contributors' perception of the current D\&I initiative, their critiques and suggestions to improve it. 

\vspace{-2pt}
\section{Research Method}
\label{sec:method}
We answer the research questions by focusing on a single large OSS foundation (ASF) to get the perspective of contributors in a single, mature OSS community instead of the broad OSS world. The ASF is the world's largest OSS Foundation with more than 460k people and more than 350 mature projects and initiatives~\cite{ASFwebsite}. The ASF serves as a good case study because of its relevancy and maturity, its interest in improving D\&I, and our collaboration with the ASF D\&I committee, who worked closely with us in the design and execution of the study, allowing us to get legitimacy in the eyes of participants and validate the interpretation of the study results.

We conducted an online survey and follow-up interviews with the ASF contributors and used mixed methods to analyze the data. 
\subsection{Survey}

\textbf{Survey design:} 
The goals of the survey were to understand ASF contributors' perception of: (1) the state of D\&I in their community and (2) the current D\&I initiative, their critiques, and ideas for improvements. The survey started with six demographic questions followed by 12 Likert-scale items and one open-ended question (see survey questions in supplemental material \cite{suppdoc}). Table \ref{tab:SurveyQuestions} presents the Likert-scale questions, from contributors' perception of role stereotyping (Q1-Q3), their ability to contribute (Q4-Q8), being represented within the community (Q9, Q10), and their perception of the code of conduct (Q11, Q12).

\begin{table*}[htb]
\caption{Twelve-Likert scale questions about participants' perception of the state of D\&I and the code of conduct. The citation indicates that a question is a replication from \citet{Lee.Carver:2019}.}
\label{tab:SurveyQuestions}
\resizebox{0.98\textwidth}{!}{
\begin{tabular}{lll}
\hline
  \textbf{\begin{tabular}[c]{@{}c@{}} Perception of \end{tabular}} &
  \textbf{Questions} &
  \textbf{Question ID} 
  \\\hline
  
  \multirow{3}{*}{Role stereotyping} &
  \cellcolor{gray!15}\begin{tabular}[l]{@{}l@{}} Other members of the project see me as a parental figure \cite{Lee.Carver:2019} \end{tabular} &
  
  \cellcolor{gray!15}\begin{tabular}[l]{@{}l@{}}Q1. Parental figure\end{tabular} \\ 
  
   &
  \begin{tabular}[l]{@{}l@{}} I am expected to take care of other members of the project more so than is usual \cite{Lee.Carver:2019}\end{tabular} &
  
  \begin{tabular}[l]{@{}l@{}}Q2. Care taker\end{tabular} \\
  
  &
  \cellcolor{gray!15}\begin{tabular}[l]{@{}l@{}} I feel some members of the community are patronizing to me  \end{tabular} & \cellcolor{gray!15}\begin{tabular}[l]{@{}l@{}}Q3. Patronized\end{tabular} \\
  \hline
 
 \multirow{5}{*}{\begin{tabular}[l]{@{}l@{}} Ability to contribute\end{tabular}} 
   &
  \begin{tabular}[l]{@{}l@{}}I have an equal chance to get contributions accepted \cite{Lee.Carver:2019} \end{tabular} &
  \begin{tabular}[l]{@{}l@{}}Q4. Equal chance\end{tabular} \\
   &
  \cellcolor{gray!15}\begin{tabular}[l]{@{}l@{}} Nothing keeps me from contributing to the project \cite{Lee.Carver:2019} \end{tabular} &
  \cellcolor{gray!15}\begin{tabular}[l]{@{}l@{}}Q5. No barriers\end{tabular} \\ 
 
   &
  \begin{tabular}[l]{@{}l@{}}I have a solid network of open-source peers \cite{Lee.Carver:2019} \end{tabular} &
  \begin{tabular}[l]{@{}l@{}}Q6. Network\end{tabular} \\
  
  &
 \cellcolor{gray!15}\begin{tabular}[l]{@{}l@{}}It was easy to find a mentor with whom I felt comfortable \cite{Lee.Carver:2019} \end{tabular} & \cellcolor{gray!15}\begin{tabular}[l]{@{}l@{}}Q7. Mentored\end{tabular} \\
 
   &
  \begin{tabular}[l]{@{}l@{}}I have a hard time following discussions because of technical jargon \end{tabular} &
 
  \begin{tabular}[l]{@{}l@{}}Q8. Tech jargon\end{tabular} \\\hline
  
 
  \multirow{2}{*}{\begin{tabular}[l]{@{}l@{}} Being represented \end{tabular}}  
  
   &
  \cellcolor{gray!15}\begin{tabular}[l]{@{}l@{}}The PMC represents a diverse set of people \end{tabular} &
 
  \cellcolor{gray!15}\begin{tabular}[l]{@{}l@{}}Q9. Diverse PMC\end{tabular} \\ 
 
   &
  \begin{tabular}[l]{@{}l@{}}I feel represented in the decision-making group \end{tabular} &
  \begin{tabular}[l]{@{}l@{}}Q10. Represented\end{tabular} \\ \hline
  

 \multirow{2}{*}{\begin{tabular}[l]{@{}l@{}} The code of conduct \end{tabular}} 
   &
    \cellcolor{gray!15}\begin{tabular}[l]{@{}l@{}}I was made aware of the code of conduct and how to report violations \end{tabular} &
 
  \cellcolor{gray!15}\begin{tabular}[l]{@{}l@{}}Q11. Aware\end{tabular} \\ 
 
   &
  \begin{tabular}[l]{@{}l@{}}I felt safer and more empowered to fully participate in this project because it followed the code of conduct  \end{tabular} &  \begin{tabular}[l]{@{}l@{}}Q12. Empowered\end{tabular}

\\ \hline 

  
\end{tabular}}
\end{table*}

We leveraged existing surveys when possible. Questions Q1, Q2, Q4-Q7 were from Lee and Carver~\cite{Lee.Carver:2019} who investigated contributor perceptions of gender in OSS. 3 of the 6 demographics questions (seniority at ASF, compensation, and gender identification) were adapted from the 2016 ASF Committer Diversity survey~\cite{2016ASFSurvey} by addressing the best practices recommended by the CHAOSS D\&I Working Group~\cite{chaossmetrics}. The other three background questions (country of residence, English confidence, and education level) were adopted from the Open Demographics Survey~\cite{opendemographicsdoc}. We used Lime Survey licensed under GPLv2, which is the world's leading open source survey software, to conduct the survey.

We then engaged with the ASF D\&I committee---composed of 18 experienced contributors with different roles including committers, Project Management Committee (PMC) members, and board members---and the community at large to refine and pilot the survey questions. An open-ended question on the state of the D\&I initiative was included based on the recommendation of the committee. 

\textbf{Survey Data Collection:}
In collaboration with the ASF D\&I committee, we invited the ASF committers to participate by sending emails to every `apache.org' email address and shared a link through the ASF developer mailing lists. Participants were first presented a consent page that described the goal of the survey, the data collection and handling policy, and who to contact (see supplemental material \cite{suppdoc}). The survey followed an opt-in strategy where participants started the survey if they agreed to voluntarily participate after reading the consent form. The survey was open for 45 days. We maintained the data confidentiality as per Apache Privacy Policies. Identifiable information or IP addresses were not collected. When participants gave email addresses for follow-up interviews, they were stored separately from the responses.

\begin{table}[tbh]
\centering
\caption{Demographics of the survey respondents (n=624).}
\label{tab:surveydemographics}
\resizebox{0.42\textwidth}{!}{
\begin{tabular}{l|r|r}
\hline
\toprule
 \textbf{Demographics} & \textbf{\#}  & \textbf{\%}  \\
 \hline
\midrule
Gender: Man                                          & 545 & 88.47\% \\
\cellcolor{gray!15}Gender: Woman                                        & \cellcolor{gray!15} 28  & \cellcolor{gray!15} 4.55\%  \\
\cellcolor{gray!15}Gender: \review{Non-binary, trans-men/women, prefer to self-describe/not-to-say}                                    & \cellcolor{gray!15}41   & \cellcolor{gray!15}6.66\%  \\

\toprule
Seniority at the ASF: Less than 1 year                      & 66  & 10.66\% \\
Seniority at the ASF: 1 to 2 years & 99  & 15.99\% \\
\cellcolor{gray!15}Seniority at the ASF: \cellcolor{gray!15}3 to 5 years              & \cellcolor{gray!15}165  & \cellcolor{gray!15}26.66\% \\
Seniority at the ASF: \cellcolor{gray!15}6 to 10 years                    & \cellcolor{gray!15}130  & \cellcolor{gray!15}21.00\% \\
\cellcolor{gray!15}Seniority at the ASF: Over 10 years                    & \cellcolor{gray!15}159  & \cellcolor{gray!15}25.69\% \\

\toprule
Education: Ph.D degree                              & 65  & 10.60\% \\
Education: Master's degree                             & 284  & 46.33\% \\
Education: Undergraduate degree                           & 200  & 32.63\% \\
\cellcolor{gray!15}Education: Technical training                   & \cellcolor{gray!15}33  & \cellcolor{gray!15}5.38\% \\
\cellcolor{gray!15}Education: High school      & \cellcolor{gray!15}30  & \cellcolor{gray!15}4.89\% \\
\cellcolor{gray!15}Education: \cellcolor{gray!15}No formal education      & \cellcolor{gray!15}1  & \cellcolor{gray!15}0.16\% \\

\toprule
English Proficiency: Very confident                          & 348  & 59.96\% \\
English Proficiency: Confident                              & 138  & 22.59\% \\
\cellcolor{gray!15}English Proficiency: Average                          & \cellcolor{gray!15}81  & \cellcolor{gray!15}13.26\% \\
\cellcolor{gray!15}English Proficiency: Uncomfortable                   & \cellcolor{gray!15}13  & \cellcolor{gray!15}2.13\% \\
\cellcolor{gray!15}English Proficiency: Not confident      & \cellcolor{gray!15}31  & \cellcolor{gray!15}5.07\% \\

\toprule
Compensation: Paid work only                              & 84  & 13.64\% \\
Compensation: A mix, but mostly paid                         & 153  & 24.84\% \\
\cellcolor{gray!15}Compensation: Unpaid only                              & \cellcolor{gray!15}247  & \cellcolor{gray!15}40.10\% \\
\cellcolor{gray!15}Compensation: A mix, but mostly unpaid                   & \cellcolor{gray!15}80  & \cellcolor{gray!15}12.99\% \\
\cellcolor{gray!15}Compensation: An equal mix of paid and unpaid      & \cellcolor{gray!15} 52  & \cellcolor{gray!15}8.44\% \\

\toprule
Continent: North America                              & 250  & 41.60\% \\
\cellcolor{gray!15}Continent: South America                                & \cellcolor{gray!15}5  & \cellcolor{gray!15}0.83\% \\
\cellcolor{gray!15}Continent: Europe                                 & \cellcolor{gray!15}237 & \cellcolor{gray!15}39.43\% \\
\cellcolor{gray!15}Continent: Africa                                 & \cellcolor{gray!15}7   & \cellcolor{gray!15}1.16\%  \\
\cellcolor{gray!15}Continent: Asia                                   & \cellcolor{gray!15}92  & \cellcolor{gray!15}15.31\% \\
\cellcolor{gray!15}Continent: Australia                                & \cellcolor{gray!15}10   & \cellcolor{gray!15}1.66\%  \\

\bottomrule

\end{tabular}}
\vspace{-1.5em}
\end{table}

We received 624 responses, resulting in a response rate of 8.5\% based on a considered total community size of 7500 contributors. We received 130 responses to our optional open-ended question about the D\&I effort at the ASF.
A majority of the 624 respondents identified as men (88.47\%) and 4.55\% as women. \review{We grouped together (see Table \ref{tab:surveydemographics}) the respondents who identified as non-binary, trans-men/women, prefer to self-describe/not-to-say to preserve their identity}. A majority of respondents were volunteers (61.53\%) and senior contributors with three or more years of experience at the ASF (73.35\%). They resided in 53 different countries located in six continents with the majority based in North America. Most of the respondents reported some level of higher education (89.56\%). 

\subsection{Interviews}

We conducted follow-up interviews to get a deeper understanding of contributors' experience at the ASF and their perspective on the D\&I initiative. From the set of 69 respondents who agreed to being contacted post-survey, we kept randomly selecting participants to interview until reaching saturation of information. We ended up interviewing 11 participants. Table \ref{table:interview participants} presents the demographics of our interview participants.
 
\begin{table*}[bht]
\caption{Interview participants' demographics.}
\resizebox{0.9\textwidth}{!}{
\begin{tabular}{lllllll}
\toprule
{\textbf{ID}} & {\textbf{Gender}} & {\textbf{Seniority at the ASF}} & {\textbf{Education}} & {\textbf{English Proficiency}} & {\textbf{Compensation type}} & {\textbf{Residence}} \\\hline
I1 & Man & 1 to 2 years & Undergraduate degree & Average & A mix, but mostly paid & Russia  \\
I2 & Man & Over 10 years & Ph.D degree & Very confident & A mix, but mostly unpaid & US  \\
I3 & Woman & 3 to 5 years & Undergraduate degree & Confident & Unpaid only & Germany  \\
I4 & Woman & Over 10 years & Undergraduate degree & Very confident & An equal mix of paid and unpaid & Ireland \\
I5 & \review{*Other} & Over 10 years & Undergraduate degree & Very confident & A mix but mostly paid  & US \\
I6 & \review{*Other} & Over 10 years & Undergraduate degree & Very confident  & Unpaid only & US  \\
I7 & Man & 3 to 5 years & Master's degree & Not confident & A mix, but mostly unpaid & Italy \\
I8 & Man & Less than 1 year & Master's degree & Very confident & Unpaid only & Japan \\
I9 & Woman & Less than 1 year & Master's degree & Confident & Unpaid only & Germany \\
I10 & Woman & 3 to 5 years & Master's degree  & Very confident & An equal mix of paid and unpaid & US    \\
I11 & Man & 3 to 5 years & Master's degree &  Very confident & An equal mix of paid and unpaid & UK \\
\bottomrule

\end{tabular}}
\label{table:interview participants}
\resizebox{0.7\textwidth}{!}{
\begin{tabularx}{\linewidth}{@{}XXX@{}}
\review{*Elided to protect identity of these respondents.}
\end{tabularx}
\vspace {-1.8em}
}

\end{table*}

Two researchers conducted the semi-structured interviews: one led the interviews while the other observed and took notes. Before each interview, we obtained the participant's consent to audio record for transcription purposes. The interview covered the participant's experience at the ASF and the mechanisms that help support the contribution process at the ASF. The interview lasted between 30 minutes to one hour, after which, we thanked our participants and, as a token of appreciation, we sent them a \$50 gift card or its equivalent in donation to the OSS project or organization of their choice. 

\subsection{Data Analysis}
We used a mixed-method approach to answer our research questions. We used ordinal logistic regression to analyze the answers to the Likert-scale questions. For the open-ended question and interview transcripts, we used open coding. 

\subsubsection{Quantitative: Ordinal Logistic Regression}
\label{sec:OLRmethod}
To analyze the Likert-scale responses, we used participants' reported demographic attributes (gender, seniority at ASF, English proficiency, compensation type, place of residence, and education level) as explanatory variables and ran ordinal logistic regression for each one of the twelve Likert-scale questions (Q1--Q12). The ordinal logistic regression model~\cite{harrell2015ordinal} is an extension of the logistic regression model where the logits of a categorical response are linearly related to the explanatory variables. 

Let $Y$ be our ordinal outcome with $j$ categories (Likert-scale values), then the ordinal logistic regression model is:

$log{\frac{P\left(Y \leq j\right)}{P\left( Y > j)\right)}}$ =  $\beta_{j_0} - \eta_1x_1 - ... - \eta_px_p$; $j$ = $1, .., j-1$

where $P\left(Y \leq j\right)$ is the cumulative probability of $Y$ less than or equal to a specific category $j$ of the response variable. The model has $j-1$ intercepts denoted by $\beta_{j_0}$ and one parameter for each explanatory variable (the demographic attributes). 

By definition, our dependent variables (Likert-scale responses) fit the first two assumptions of the ordinal logistic regression model: The dependent variables are ordered and one or more of the explanatory variables are either continuous, categorical or ordinal. We checked for the absence of multi-collinearity by generating a covariance matrix of our variables. The last assumption is the proportional odds which ensures that the relationship between each pair of outcome groups is the same, meaning that there is only one set of coefficients, which means that there is only one model. We used the Brant test to check for the proportional odds assumption. We conclude that the parallel assumption holds when the probability (p-values) for all variables are greater than $\alpha$ = $0.05$ and the Omnibus variable, which stands for the whole model, is also greater than $\alpha$. In the case where an explanatory variable failed the Brant test, we omitted that variable from the model (see grayed out cell in Tables \ref{tab:genderStereotypingOLR}, \ref{tab:abilityToContributeOLR} and \ref{tab:beingRepresentedOLR}) to ensure that the model fits all the assumptions. We used R version 4.0.4 and $polr$ in the ``MASS'' package for the analyses.

Note, for the purpose of this analysis, we divided each demographic attribute (e.g., seniority at ASF, see Table \ref{tab:surveydemographics}) into two segments. For instance, we grouped ASF contributors with low experience (<1 year, 1-2 years) as junior contributors and those with more experience (3-5, 6-10 and >10 years) as senior contributors. \review{We also grouped the participants according to their gender: ``gender-majority'' for the ASF contributors who identified as men, and ``gender-minority'' for the ASF contributors who identified as women, non-binary, trans-men/women, prefer to self-describe/not-to-say}.

\subsubsection{Qualitative: Open Coding}
We used open coding to analyze our open-ended question and interview data. First, we analyzed the survey open-ended question on the state of D\&I (see survey questions in supplemental \cite{suppdoc}). We inductively coded the answers, built post-formed codes as the analysis progressed and associated them to respective parts of the text. At this stage, our aim was to code contributors' perception of D\&I according to their discourse, and not according to any preconceived data. For each interviewee, we identified and coded each excerpt that presented a perception of the D\&I. 
Once a week, the research team met to discuss the emerging categories and to refine their nomenclature. The coding process was conducted by one researcher and discussed with the other researchers until consensus about the resulting codes and quotes. 
\section{Results}
In the following, we present the analysis of how contributors perceive (1) the state of D\&I in their community (RQ1, Section 4.1) and (2) the current D\&I efforts (RQ2, Section 4.2). 

\subsection{Perceptions on the state of D\&I}
\label{sec:RQ1}

To answer RQ1, we analyzed the survey and interviews using mixed methods. Sections \ref{sec:role_streo} to \ref{sec:being_represented} detail contributors' perceptions of different aspects of the state of D\&I (see Table \ref{tab:SurveyQuestions}), and Section \ref{sec:extentIssues} illustrates contributors' perceptions of the extent of the D\&I issues. We segment the analysis based on the demographic attributes.

Tables \ref{tab:genderStereotypingOLR}, \ref{tab:abilityToContributeOLR}, and \ref{tab:beingRepresentedOLR} show the ordinal logistic regression results for the survey questions. For instance, the model ``Q2. Care taker'' in Figure \ref{fig:likert_role_stereotyping} shows how the demographic attributes explain the level of agreement on that question (see Section~\ref{sec:OLRmethod}). The tables highlight the variables that show statistical significance at $p<.05$ or $p<.01$ post fixed by \textsuperscript{$*$} or \textsuperscript{$**$}, respectively. 
We also show the response distributions according to the Likert-scale for those demographic attributes that were statistically significant in our models (Figures ~\ref{fig:likert_role_stereotyping}, \ref{fig:likert_ability_contribute}, \ref{fig:likert_being_representation}).
In the rest of this section, we discuss \textit{three} different aspects of the contributors' perceptions, namely \textit{role stereotyping}, contributors' \textit{ability to contribute}, and their \textit{feeling of being represented}. 

\subsubsection{Role Stereotyping} \label{sec:role_streo}
\begin{table*}[htb]
\centering

\caption{Ordinal logistic regression for role stereotyping Likert-scale questions. The \colorbox{green!20}{highlight} corresponds to a statistically significant difference. OR refers to the odds ratio calculated as the exponential of the ordinal logistic regression value.}
\label{tab:genderStereotypingOLR}
\resizebox{0.63\textwidth}{!}{%
\footnotesize
\begin{tabular}{l|l|l|l|l|l|l}
\hline
\toprule
\multicolumn{1}{l|}{\textbf{Background attributes}} &
\multicolumn{2}{l|}{\textbf{Q1. Parental figure}} &
\multicolumn{2}{l|}{\textbf{Q2. Care taker}} &
\multicolumn{2}{l}{\textbf{Q3. Patronized}} 
\\ 
\multicolumn{1}{l|}{}&
OR  &
Std. err &
OR  &
Std. err &
OR &
Std. err
\\ \hline
\midrule
 
\review{Gender-minority} vs. \review{Gender-majority}  & 
1.09  & 
0.29 & 
1.75 &
0.29 &
\cellcolor{green!20} 3.36** &
\cellcolor{green!20} 0.29
\\
\toprule

Senior vs. Junior & 
\cellcolor{green!20}2.48** & 
\cellcolor{green!20}0.21 &
\cellcolor{green!20}1.86** &
\cellcolor{green!20}0.20 &
0.86 &
0.22
\\
\toprule

English non-confident  vs. confident  & 
0.79 &
0.23 &
\cellcolor{green!20}1.67* &
\cellcolor{green!20}0.21 &
\cellcolor{green!20}2.38** &
\cellcolor{green!20}0.23
\\
\toprule

College vs. No college &
0.81 &
0.29 &
0.70 &
0.27 &
0.91 &
0.32
\\
\toprule

Paid vs. Non-paid & 
1.02 &
0.18 &
1.11 &
0.18 &
0.81 &
0.21
\\
\toprule

North America vs. Outside  & 
0.81 & 
0.19 &
\cellcolor{gray!22} &
\cellcolor{gray!22} &
\cellcolor{gray!22} &
\cellcolor{gray!22}  
\\


\bottomrule
\end{tabular}}
\resizebox{0.7\textwidth}{!}{
\begin{tabularx}{\linewidth}{@{}XXX@{}}
Odds  ratio (OR)  greater  than  1  means  that  the  first  segment  has greater  chances  of agreeing with the question than the second. Ratio  less  than  1  means  the  opposite.
We tested the proportional odds assumption for each question using the Brant test. The grayed out cells correspond to the explainable variables that are omitted from the model to satisfy the proportional odds assumption (see section \ref{sec:OLRmethod}). A single * means that the variable is significant at $p<.05$ and a ** means that a variable is significant at  $p<.01$.
\end{tabularx}
\vspace {-1.8em}
}
\end{table*}

\begin{figure}[hbtp]
    \centering
    \includegraphics[width=.48\textwidth]{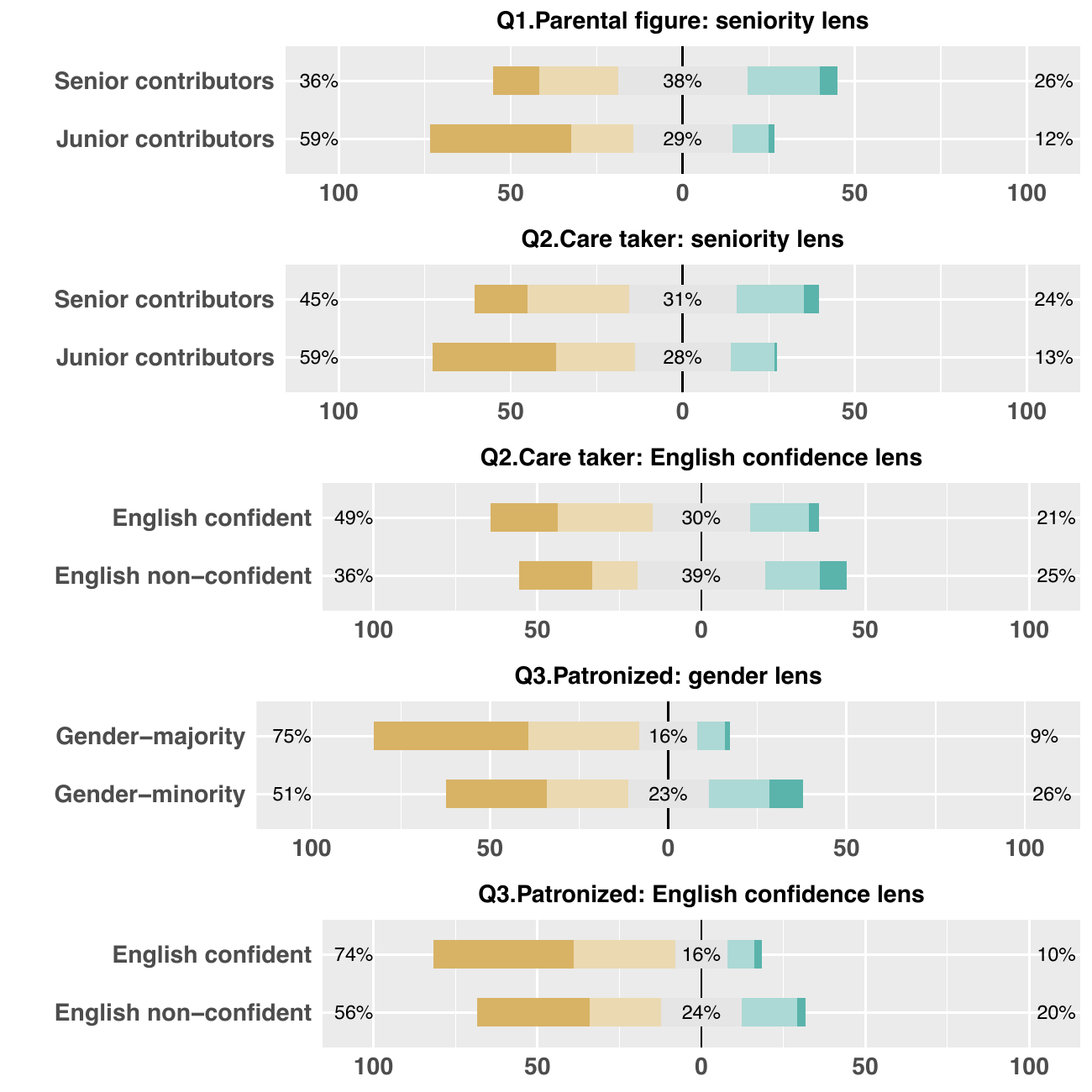}
    \caption{Responses to the 5-points Likert-scale items about role stereotyping (Q1-Q3, Table \ref{tab:SurveyQuestions}). Left hand (yellow) shows levels of disagreement, middle (grey) is neutral, and right (green) shows level of agreement. We only show the responses where the differences between segments were significant. See supplemental material \cite{suppdoc} for all the Likert-scale figures.}
    \label{fig:likert_role_stereotyping}
    \vspace*{-1.93em}
\end{figure}

Past work has shown that gender can be a major source of bias in how people perceive others, which is linked to prescribing certain roles and traits to women~\cite{wyer2014advances, denmark2007psychology}, and feminine attributes or qualities displayed by women tending to be devalued~\cite{barreto2015detecting, fiske2015intergroup}. Gender is also closely linked to two basic dimensions that individuals rely on to judge other people---when an individual meets someone, they intuitively make judgments of their warmth and competence~\cite{cuddy2008warmth, fiske2007universal}, subjecting women to role stereotyping. Women are stereotyped to be motherly, warm, and nurturing~\cite{mckinnon2020perceptions, carli2016stereotypes, heilman2007women, mitchell2018gender}. Such stereotypes perpetuate behavioral expectations for women to assume the role of the parental figure and community care taker~\cite{mckinnon2020perceptions, kaplan1994woman}. 

Questions Q1-Q3 explore three types of role stereotyping, namely \textsc{Parental figure} (Q1), \textsc{Care taker} (Q2), and being \textsc{Patronized} (Q3), and the extent to which individuals from different backgrounds perceive role stereotyping to take place at the ASF. 
 
While women have been associated with the role stereotype of \textsc{Parental figure}, we found that contributors \review{who identified as gender-majority} and \review{those who associate with the gender-minority} reported at similar rates being seen as a parental figure (Q1 in Table \ref{tab:genderStereotypingOLR}). This is inline with \citet{Lee.Carver:2019}'s results. On the other hand, looking at different attributes, we found that senior ASF contributors, were 2.48 times more likely than juniors to feel pigeonholed into this role (Table~\ref{tab:genderStereotypingOLR}, $p<.01$). This difference can be seen in the distribution of responses in Figure~\ref{fig:likert_role_stereotyping}---senior contributors agreed at a higher rate with the statement as compared to junior contributors, who disagreed at a higher rate. The results indicate that those more experienced with the ASF are being sought out for advice and mentoring.

Women have also been associated with \textsc{care taker} roles. Our results for this question (Q2 in Table \ref{tab:SurveyQuestions}) were similar to the aforementioned results. While the odds of  \review{contributors who associate with gender-minority} feeling that they were expected to take care of others were 1.75 times higher than their counterparts, this difference was not statistically significant. The seniority lens, however, painted a different picture in which the odds of senior contributors feeling expected to take care of others was 1.86 times ($p<.01$) higher than that of juniors (see Table \ref{tab:genderStereotypingOLR}, Figure \ref{fig:likert_role_stereotyping}). Non-confident English speakers were also more likely to report being expected to take on \textsc{Care taker} roles (1.67 times more than confident English speakers with $p<.05$, see Table \ref{tab:genderStereotypingOLR}). A possible explanation is that non-confident English speakers might have overcome language barriers when contributing to OSS and are sought out by others who face similar challenges. 

Finally, we asked whether participants felt \textsc{Patronized} by other community members (Q3 in Table \ref{tab:SurveyQuestions}). \review{Contributors who identified as gender-minority and who reported being} non-confident English speakers felt to be so with odds ratios 3.36 ($p<.01$) and 2.38 ($p<.01$), respectively. This suggests that these minorities do not feel equally respected as their counterparts. The literature reports that those in the minority have to prove themselves and may not be taken seriously~\cite{Lee.Carver:2019}, perpetuating the social barriers they face~\cite{steinmacher2015social}.

\MyBox{\textbf{Role Stereotyping.} Gender stereotypes of \textsc{Parental figure} or \textsc{Care taker} were not significant factors for respondents who identify as \review{gender-minority}, but feeling \textsc{Patronized} was. Senior contributors and those not confident in English, however, did feel stereotyped as being parental figures or care givers.} 
\vspace{-0.8cm}

\subsubsection{Perceptions about ability to contribute}
\label{sec:ability_contribute}
\begin{table*}[htb]
\centering
\caption{Ordinal logistic regression for ability to contribute. 
}
\label{tab:abilityToContributeOLR}
\resizebox{0.82\textwidth}{!}{
\begin{tabular}{l|l|l|l|l|l|l|l|l|l|l}
\hline
\toprule
\multicolumn{1}{l|}{\textbf{Background attributes}} &
\multicolumn{2}{l|}{\textbf{Q4. Equal chance}} & \multicolumn{2}{l|}{\textbf{Q5. No barriers}} &
\multicolumn{2}{l|}{\textbf{Q6. Network}} & 
\multicolumn{2}{l|}{\textbf{Q7. Mentored}} &
\multicolumn{2}{l}{\textbf{Q8. Tech jargon}} 
\\ 
\multicolumn{1}{l|}{}&
OR  &  Std. err & OR  & Std. err & OR & Std. err & OR & Std. err & OR & Std. err \\ \hline
\midrule
 
\review{Gender-minority} vs. \review{Gender-majority} & 
0.59  & 
0.34 & 
\cellcolor{green!20} 0.41** &
\cellcolor{green!20} 0.29 &
\cellcolor{green!20} 0.52** &
\cellcolor{green!20} 0.29 &
\cellcolor{gray!22} &
\cellcolor{gray!22} &
1.54 &
0.35
\\
\toprule

Senior vs. Junior  & 
1.31 & 
0.24 &
1.40 &
0.22 &
\cellcolor{green!20} 2.07** &
\cellcolor{green!20} 0.20 &
0.96 &
0.21 &
\cellcolor{gray!22} &
\cellcolor{gray!22} 
\\
\toprule

English non-confident vs. Confident & 
1.12 &
0.30 &
1.01 &
0.25 &
\cellcolor{green!20} 0.48** &
\cellcolor{green!20} 0.24 &
\cellcolor{gray!22} &
\cellcolor{gray!22} &
\cellcolor{green!20} 2.58** &
\cellcolor{green!20} 0.28
\\

\toprule

College vs. No college &
0.44 &
0.49 &
0.81 &
0.36 &
1.15 &
0.31 &
1.28 &
0.32 &
0.91 &
0.37
\\

\toprule

Paid vs. Non-paid & 
0.78 &
0.23 &
1.01 &
0.21 &
0.85 &
0.20 &
0.87 &
0.19 &
0.70 &
0.25
\\

\toprule

North America vs. Outside & 
1.12 & 
0.24 &
\cellcolor{gray!22} &
\cellcolor{gray!22} &
1.13 &
0.21 &
\cellcolor{gray!22} &
\cellcolor{gray!22} &
1.25 &
0.27
\\


\bottomrule
\end{tabular}}

\end{table*}

\begin{figure}[hbtp]
    \centering
    \includegraphics[width=.48\textwidth]{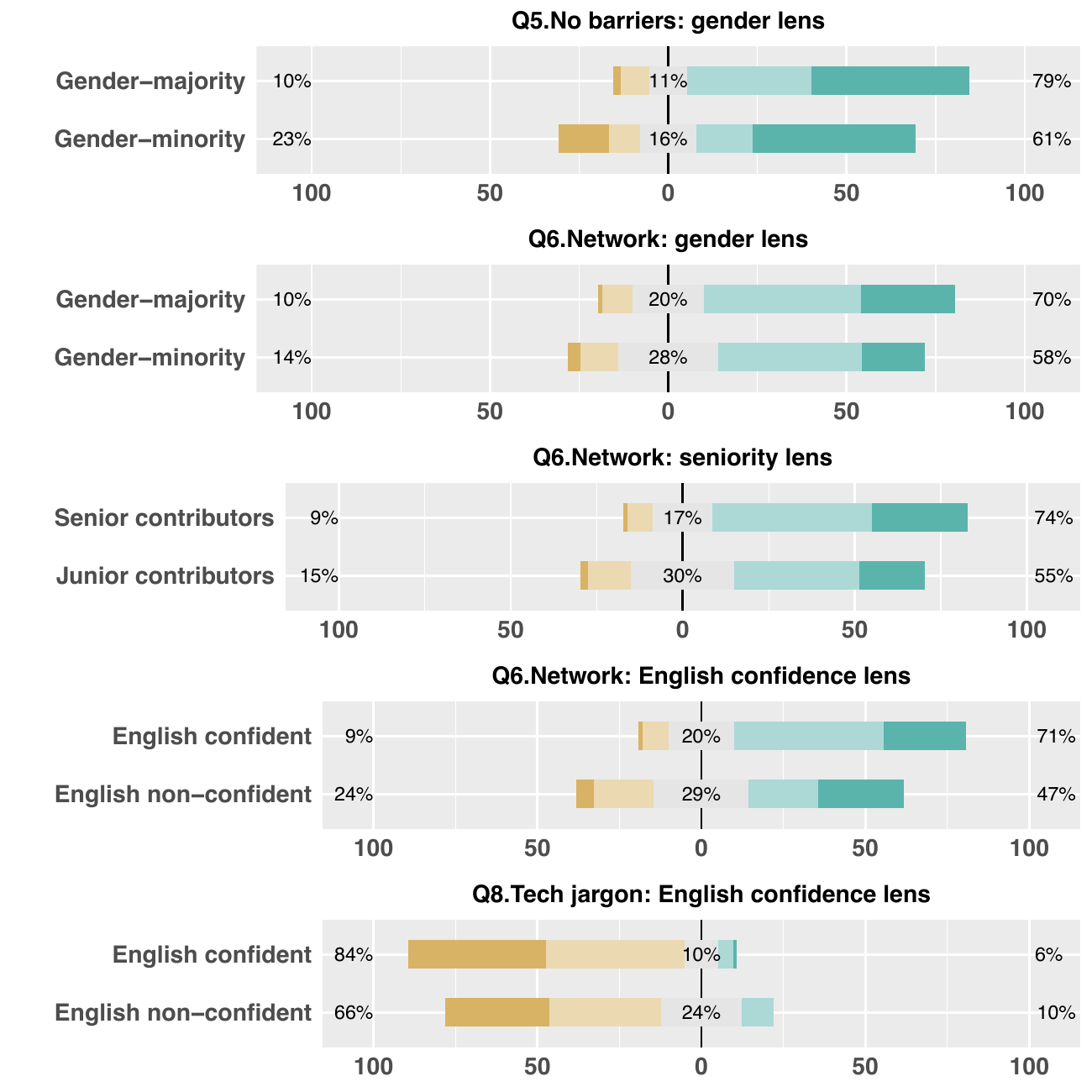}
    \caption{Responses to the 5-points Likert-scale items for `ability to contribute' (Q5, Q6, Q8 were significant).}
    \label{fig:likert_ability_contribute}
    \vspace{-1.45 em}
\end{figure}

Past studies have reported that the ``so called'' meritocracy~\cite{nafus2012patches} in OSS is biased and include hostile environments~\cite{feller2000framework} that create barriers to contribution~\cite{steinmacher2015systematic,balali2018newcomers,steinmacher2015social, vasilescu2015perceptions}. \citet{Lee.Carver:2019} recently reported the perceptions of women contributors of their ability to contribute to OSS and the barriers they faced. Therefore, we aimed to understand how contributors with different backgrounds perceive their ability to contribute (questions Q4-Q8 in Table~\ref{tab:abilityToContributeOLR}). 

We found that none of the demographic attributes make a significant difference in participants' perceptions of having an \textsc{Equal chance} (Q4) to contribute and in finding a mentor (\textsc{Mentored} (Q7)). However, a more nuanced story emerges when considering the responses to questions regarding \textsc{No barriers} to contributing (Q5), having a solid \textsc{Network} (Q6), and challenges in following discussions because of \textsc{Tech jargon} (Q8) (see Table \ref{tab:abilityToContributeOLR}).

\review{Contributors who associate with the gender-minority group} were less likely to agree (0.41) that there were \textsc{No barriers} (Q5 in Table~\ref{tab:SurveyQuestions}). That means that they were 2.44 times more likely to agree that they faced barriers ($\frac{1}{0.41}$ = $2.44$). These findings align with prior literature~\cite{Terell-2017, mendez2018open}. 

With regards to \textsc{Network}, the respondents showed mixed feelings depending on the demographic attribute. As Figure~\ref{fig:likert_ability_contribute} shows, 74\% of senior ASF contributors reported having a solid network of OSS peers with the odds 2.07 times ($p<.01$) higher than their junior counterparts (Table~\ref{tab:abilityToContributeOLR}). This can probably be explained by senior contributors' tenure and the connections built over the years, along with the fact that OSS communities can be hard for junior contributors to join~\cite{steinmacher2015social}. 

While those who identified as \review{gender-majority} were more likely to report (1.92 times) having a solid network of peers ($p<.01$), only 58\% of \review{gender-minority} contributors agreed to the statement (Figure~\ref{fig:likert_ability_contribute}). This is worrisome since prior literature emphasizes the importance of same gender role models, peers in building a network, and sense of belonging~\cite{trinkenreich2020hidden,calvo2020,powell2010}.

While building a solid network can be challenging for a variety of reasons~\cite{steinmacher2015social, storey2016social}, contributors who are not confident with their English speaking skills bear an additional communication burden. In fact, our results show that contributors who are not confident in English were 2.58 times more likely ($p<.01$) to report struggling to follow discussions because of technical jargon (see Table~\ref{tab:abilityToContributeOLR}). The language barrier may impact contributors' participation in communication channels, thus limiting their chances to build meaningful connections with their peers. Contributors who were confident English speakers were 2.08 times (Table~\ref{tab:abilityToContributeOLR}, $p<.01$) more likely to report having a solid network of OSS peers.

\MyBox{\textbf{Ability to contribute.} Gender \review{was relevant in the perception of barriers} that keeps those in \review{the gender-minority group} from contributing or building a solid network of peers. Lack of proficiency in English is also associated with barriers in creating a network or following technical discussions that include jargon.} 
\subsubsection{Perceptions of being represented}
\label{sec:being_represented}

The feeling of being represented is important to being productive and satisfied~\cite{lim2008job}.
\citet{hagerty1996sense} emphasize the importance of the experience of personal involvement in a system or environment, so that people feel as an integral part of it. Past work found that certain factors can impede the feeling of being represented, such as the lack of interpersonal relationships in the community~\cite{hoyle1994use, tinto1987leaving} and the perception that one's voice is lost in an environment where the loudest voice prevails~\cite{nafus2012patches}.

\begin{table}[htb]
\centering
\caption{Ordinal logistic regression for being represented.
}
\label{tab:beingRepresentedOLR}
\resizebox{0.48\textwidth}{!}{
\begin{tabular}{l|l|l|l|l}
\hline
\toprule
\multicolumn{1}{l|}{\textbf{Background attributes}} &
\multicolumn{2}{l|}{\textbf{Q9. Diverse PMC}} &
\multicolumn{2}{l}{\textbf{Q10. Represented}}
\\ 
\multicolumn{1}{l|}{}&
OR  &
Std. err &
OR  & Std. err 
\\ \hline
\midrule
 
\review{Gender-minority} vs. \review{Gender-majority} & 
0.68  & 
0.29 & 
\cellcolor{gray!22} &
\cellcolor{gray!22} 
\\

\toprule

Senior vs. Junior  & 
0.86 & 
0.20 &
\cellcolor{green!20} 1.67* &
\cellcolor{green!20} 0.22
\\

\toprule

English non-confident vs. Confident & 
0.97 &
0.22 &
\cellcolor{green!20} 0.60** &
\cellcolor{green!20} 0.25
\\

\toprule

College vs. No college &
0.99 &
0.28 &
0.76 &
0.36
\\

\toprule

Paid vs. Non-paid & 
\cellcolor{green!20} 0.52** &
\cellcolor{green!20} 0.18 &
0.93 &
0.22 
\\

\toprule

North America vs. Outside  & 
0.87 & 
0.19 &
1.35 &
0.24
\\


\bottomrule
\end{tabular}}
\end{table}
\begin{figure}[hbtp]
    \centering
    \includegraphics[width=.48\textwidth]{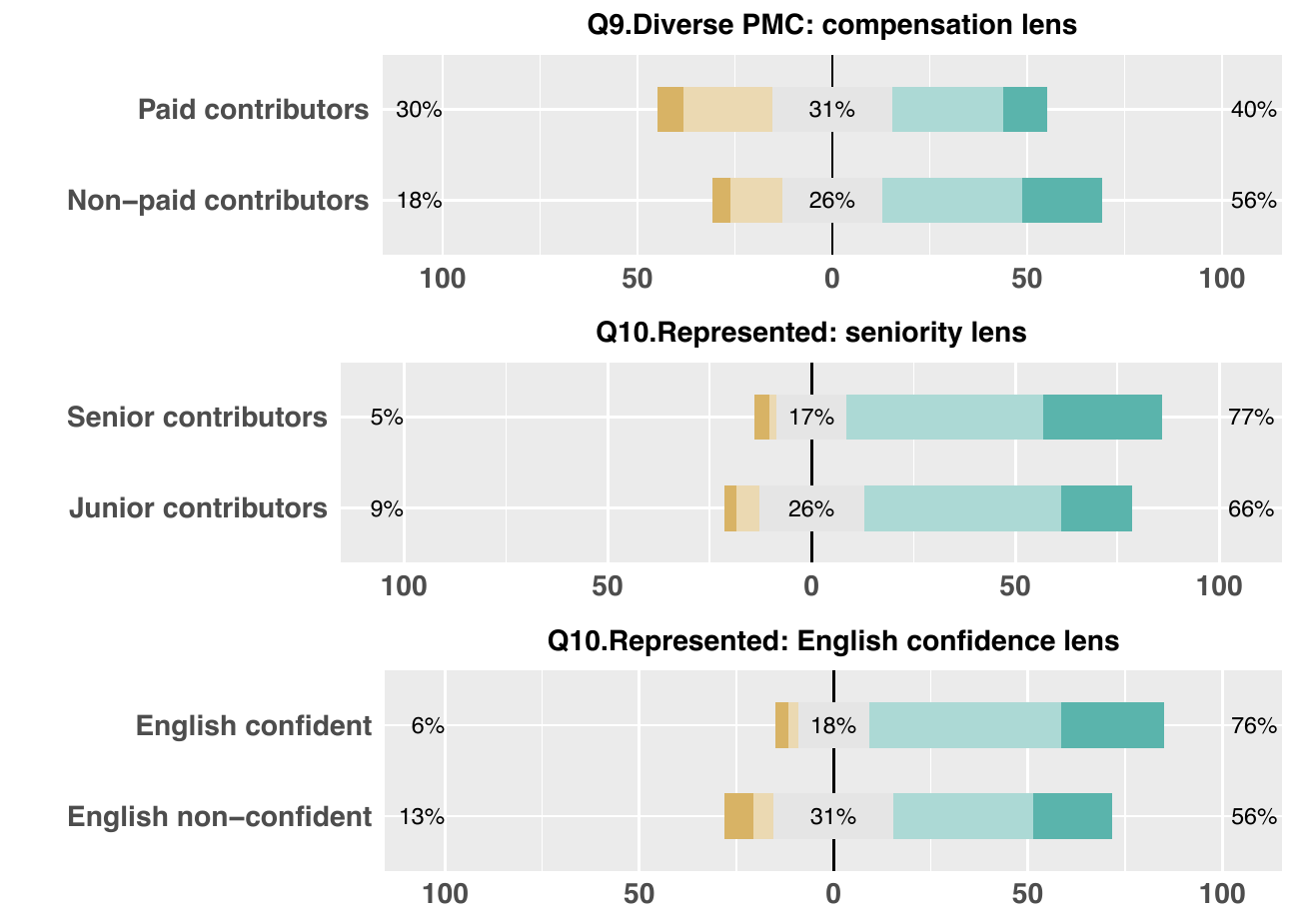}
    \caption{Responses to the 5-points Likert-scale items for `being represented' (Q9 and Q10 were significant).}
    \label{fig:likert_being_representation}
    \vspace{-1.45em}
\end{figure}

We asked contributors to rank their agreement on whether they felt that ASF has a \textsc{Diverse PMC} (Q9) and whether they felt being \textsc{Represented} (Q10) (see Table \ref{tab:SurveyQuestions}).

For \textsc{Diverse PMC}, we only found statistically significant difference when analyzing paid vs. non-paid contributors. More than half of volunteer contributors reported that the PMC was diverse as compared to 40\% of the paid contributors (see Figure \ref{fig:likert_being_representation}). Indeed, as Table \ref{tab:beingRepresentedOLR} shows, the odds of non-paid contributors agreeing to \textsc{Diverse PMC} statement were 1.92 times higher than that of their paid peers ($p<.01$).

This highlights a possible tension in the relatively new hybrid-model of OSS, where paid employees making contributions do not perceive that the PMC is diverse enough. Since none of the other factors played a significant role, we assume here that the paid contributors did not see themselves being represented in the PMC.  

Regarding the question about feeling \textsc{Represented} in the decision making process, the seniority and the English confidence lens revealed differences. Senior ASF contributors were 1.67 times more likely ($p<.05$) to agree that they feel represented in decision making processes. Similarly, contributors who are confident in their English speaking skills were 1.67 times more likely ($p<.01$) to agree than their counterparts (Table~\ref{tab:beingRepresentedOLR}). In fact, 76\% of participants who were confident in speaking English agreed to \textsc{Represented} as compared to 56\% of those who were not confident in English. These findings are in line with past work that has found language and nationality to be a barrier in OSS~\cite{prana2020including, ortu2017diverse, robles2016women}.  

\MyBox{\textbf{Being represented.} English speaking skills, compensation, and seniority attributes were relevant in participants' perceptions of feeling represented in the PMC or decision making.}
\subsubsection{Perceptions of extent of D\&I Issues}
\label{sec:extentIssues}

\begin{figure}[hbtp]
    \centering
    \includegraphics[width=.48\textwidth]{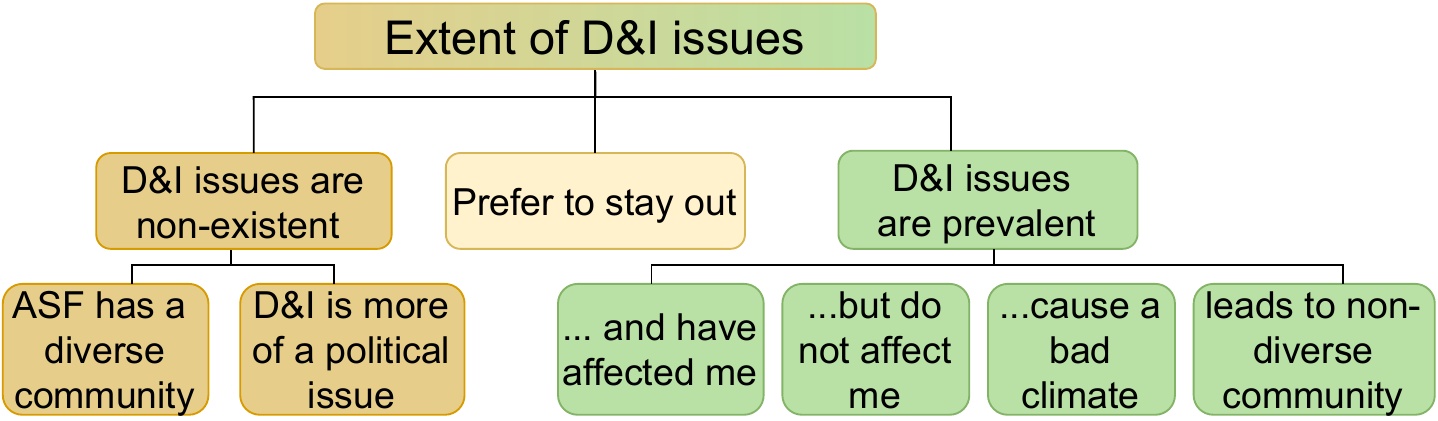}
    \caption{ASF contributors' perspective on the extent of D\&I issues.}
    \label{fig:rq1_extentOfD&I}
    \vspace{-1.5em}
\end{figure}

To investigate contributors' perceptions of D\&I issues at the ASF, we qualitatively analyzed the open-ended survey question and interviews. Figure \ref{fig:rq1_extentOfD&I} presents a summary of the results, showing the contrasting perceptions of the extent of the D\&I issues. 

At one end of the spectrum, some contributors reported \textsc{D\&I issues are non-existent} and that the community was diverse. For instance, S795 reported that the ``ASF consists of maximum diversity'' and that ``diversity of physical attributes are both invisible and largely ignored [in] mailing lists.'' Similarly, S36 said that ``diversity has rarely been an actual issue'' and that it is ``used for political pressure that compromises free collaboration.'' These responses reflect the belief that some feel D\&I is an unsubstantiated, political issue. 

Some preferred to \textsc{stay out} of discussing D\&I, such as S517, who shared ``I have therefore withdrawn my active participation in all of these types of usually quite `loud' discussions.'' 

On the other end of the spectrum, contributors reported \textsc{D\&I issues are prevalent}, because of their first-hand experiences or that of others. 
Multiple participants who were in gender minority groups reported being affected by a biased committer selection process, as I5 shared: ``it took a very, very long time for me to become a committer relative to other people who became committers...it's also really, really hard to add other non-men as committers.'' I3 (woman) had a similar experience explaining how it was hard for non-code contributors to become committers. 

Some contributors described the community as an``old boys club'' (S618) and reported noticing D\&I issues. For instance, S259 reported: ``I never experienced real negative emotions towards me, but I have seen people saying stupid things''.
Some of them felt that the ASF was not a diverse community, when considering gender: ``Of the 54 committers and 33 PMC members of [project name], just one (that I am aware of) identifies as a woman''(S753); or race: ``the number of African American PMC members is even more unrepresentative of the US population''(S821).

\MyBox{\textbf{Extent of D\&I issues.} Contributors' perceptions of D\&I issues are a mixed bag. Some attest to its presence from their or others' experiences. However, there are others who feel that D\&I issues are non existent and politicized.}
\subsection{Perceptions on D\&I initiative at the ASF}
\label{sec:RQ2}

In this section, we answer RQ2 by discussing participants' perceptions of the ASF D\&I initiative. We start by looking at how they perceive the code of conduct specifically; then, we analyze their general perceptions of the ASF D\&I initiative. 

\subsubsection{Code of conduct}
A Code of Conduct (CoC) defines the expected behavior for the project's community, which, when adopted, can help foster a positive social atmosphere~\cite{Coc}. Multiple communities, such as Django, Python, Ubuntu, Contributor Covenant, and Geek Feminism, adopted the code of conduct early on and contributed to its adoption in other communities~\cite{tourani2017}. It has now become one of the most adopted D\&I effort in OSS~\cite{li2021code}. The goal of questions \textsc{Q11. Aware} and \textsc{Q12. Empowered} (see Table~\ref{tab:SurveyQuestions}) was to evaluate the perceptions of contributors about the ASF CoC. 

Therefore, we asked the contributors whether they were aware of the ASF CoC and ways to apply it. There were no significant differences between segments across the different demographic attributes. The majority of contributors indicated that they were aware of the CoC (see Likert-scale figures in supplemental~\cite{suppdoc}). This is probably because CoCs are now widely adopted in most OSS communities. 

\begin{figure}[hbtp]
    \centering
    \includegraphics[width=.48\textwidth]{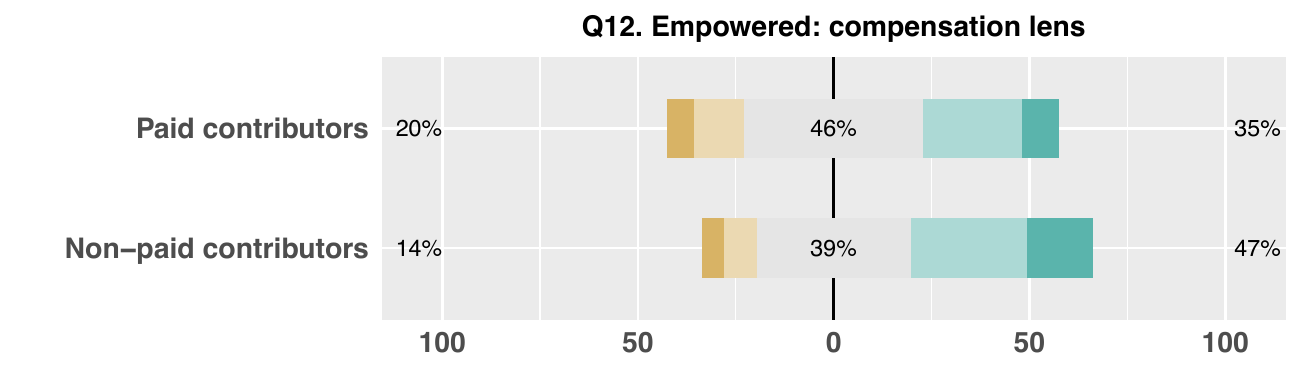}
    \caption{Responses to the 5-points Likert-scale item for `code of conduct' (Q12 was significant).}
   \vspace{-0.9em}
    \label{fig:likert_CoC}
\end{figure}
\begin{table}[htb]
\centering
\caption{Ordinal logistic regression for the code of conduct. 
}
\label{tab:CoCOLR}
\resizebox{0.47\textwidth}{!}{
\begin{tabular}{l|l|l|l|l}
\hline
\toprule
\multicolumn{1}{l|}{\textbf{Background attributes}} &
\multicolumn{2}{l|}{\textbf{Q11. Aware}} &
\multicolumn{2}{l}{\textbf{Q12. Empowered}} 
\\ 
\multicolumn{1}{l|}{}&
OR  &
Std. err &
OR  &
Std. err
\\ \hline
\midrule
 
\review{Gender-minority} vs. \review{Gender-majority} & 
0.89  & 
0.29 & 
0.88 &
0.30
\\

\toprule

Senior vs. Junior  & 
0.87 & 
0.19 &
0.89 &
0.20
\\

\toprule

English non-confident vs. Confident & 
0.83 &
0.22 &
0.97 &
0.23 
\\

\toprule

College vs. No college &
1.27 &
0.27 &
1.29 &
0.29
\\

\toprule

Paid vs. Non-paid & 
0.70 &
0.18 &
\cellcolor{green!20} 0.60** &
\cellcolor{green!20} 0.19
\\

\toprule

North America vs. Outside & 
0.88 & 
0.19 &
1.02 &
0.20 
\\


\bottomrule
\end{tabular}}
\vspace*{-1em}
\end{table}

We then followed up by asking contributors whether the presence of the CoC helped them feel safer and more \textsc{Empowered} within their community. The compensation lens (paid vs. non-paid) was the only  attribute that showed a statistically significant difference. 47\% of non-paid contributors felt safer and more empowered because of CoC, as compared to 35\% of paid contributors (see Figure~\ref{fig:likert_CoC}). In fact, in our survey, non-paid contributors were 1.67 times more likely ($p<.01$) to agree that CoC made them feel empowered (Table \ref{tab:CoCOLR}).

\MyBox{\textbf{Code of conduct.} Participants are aware of the CoC and non-paid contributors reported being empowered by it.}

\subsubsection{D\&I initiative at the ASF}
\label{sec:feedback}
As OSS communities have become aware of inclusivity problems, they have created different D\&I initiatives and committees~\cite{EDIgroup}. While this is a great start, it can backfire if the efforts are fragmented or incorrectly implemented. It is thus crucial to understand, early on, how contributors perceive these initiatives to ensure their success. 

The open-ended survey question was focused on collecting contributors' thoughts about the D\&I initiative at the ASF. This question was answered by 130 (out of 624) participants---\review{105 identified as men (gender-majority) and 25 were categorized as gender-minority.}
We triangulated the survey results with the interview results.
\begin{figure}[hbtp]
    \centering
    \includegraphics[width=.48\textwidth]{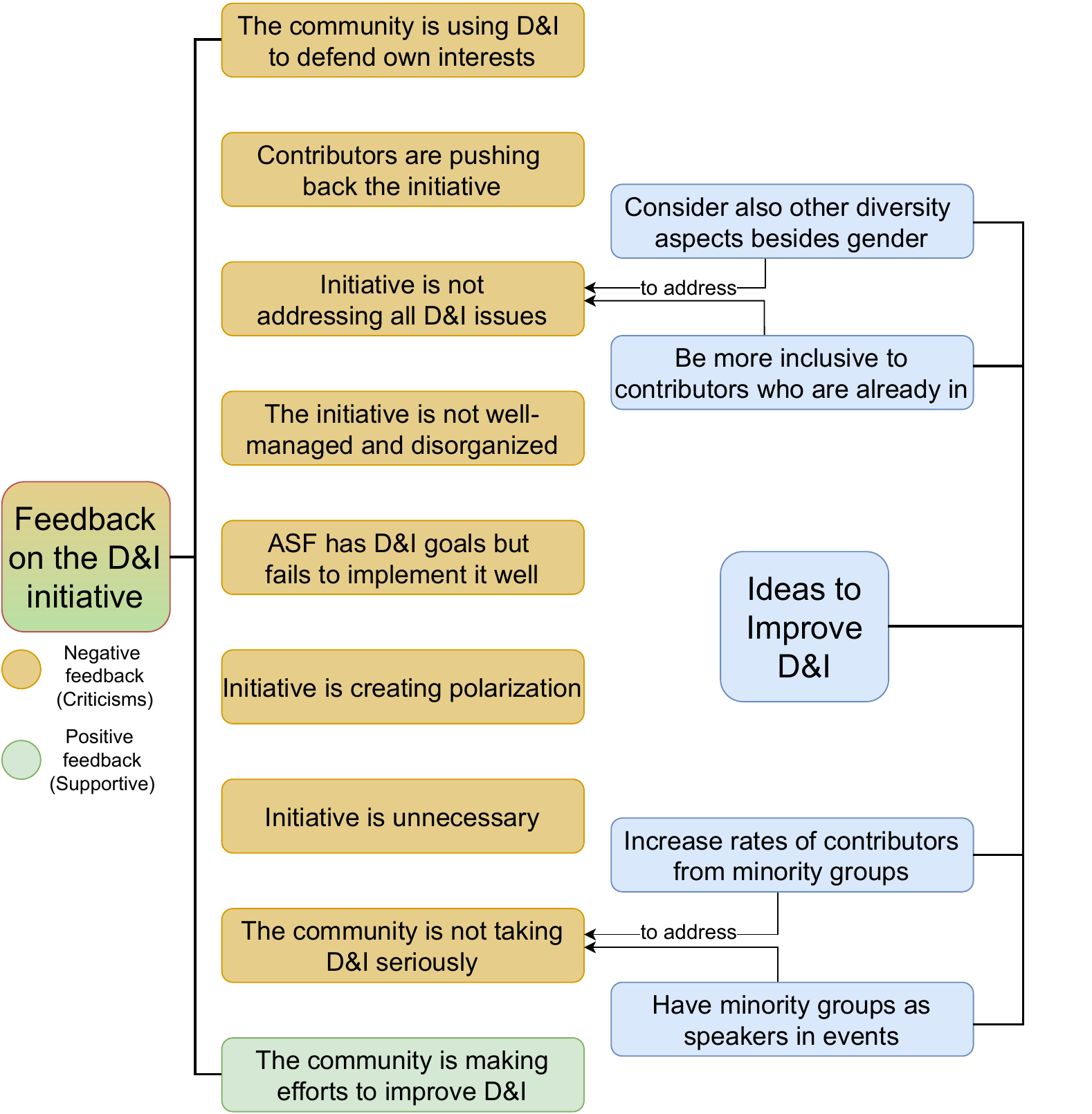}
    \caption{The feedback and ideas from ASF contributors regarding the D\&I initiative.}
    \label{fig:rq2}
    \vspace{-0.8em}
\end{figure}
Our analysis revealed 9 categories about their \textit{feedback} on the D\&I initiative and 4 categories of \textit{ways to improve} this initiative. We built a perception model backed by the data, composed of these two perspectives (see Figure \ref{fig:rq2}).

\textbf{Feedback on current D\&I initiative.}
Participants reported eight negative (or criticisms) and one positive (or supportive) perceptions. 

The participants who provided negative feedback considered that the ASF and contributors are not aligned with the same goal. On one side, participants reported that \textit{the community is using D\&I to defend their own interest}, which can be a distortion manifested by either political or personal announcements---``using the cloak of diversity to fight their own fights'' (S517). On the other side, participants voiced frustrations, raising that \textit{contributors are pushing back the initiative} and ``fighting to block D\&I efforts at the ASF'' (S38).

One possible reason for the lack of contributors' support can be that \textit{the D\&I initiative is not addressing broader issues}, as ``gender is only a tiny part of the picture'' (S340). Participants felt that the ASF is applying a limited view of the many diversity aspects by considering only gender and dismissing ``other factors that one would assume goes into the make-up of diversity'' (S181). Another reason keeping contributors from being supportive is when they consider that \textit{the D\&I initiative is not well-managed and organized}. One participant reported that ``nobody attends events more than once a week" and that the event frequency options were unrealistic, even for a person who ``attends events more than almost anyone" (S340). Other participants mentioned that discussions occur in general mailing lists  without adequate moderation, which makes these discussions ``noisy, less focused, non-constructive, emotionally driven, and empathy-lacking" (S52, S353, S640). Thus, disorganization can be a problem if \textit{the ASF has D\&I goals, but fails to implement it well}. 

A consequence of a lack of support from stakeholders is that \textit{the initiative is creating polarization} and may ``cause divisions" (S239) between contributors in minority groups and those who are not. S853 reported feeling ``oppressed by sheer volume and intolerance in D\&I movements in ASF...where anyone not in minority currently under protection is ganged upon and perceived as inherently inferior...feel unwelcome despite being caucasian heterosexual male". 

We found contradiction about the value of this initiative. Some contributors stated that \textit{the initiative is unnecessary} or ``irrelevant to ASF" (S745). This was echoed by other respondents saying that D\&I is ``getting way too much attention" (S492), and the ASF should not ``spend over much time on D\&I" (S392). Meanwhile, other participants considered that \textit{the community is not taking D\&I seriously} (S32, S703, S843, I10) and is not dedicating ``a lot of momentum around the initiatives" (I10). This points that more effort is needed in building awareness in the broader ASF community about D\&I issues and their impact on minority groups, as well as mechanisms to create allies from the majority groups.
However, some participants provided positive feedback stating that \textit{the community is making efforts to improve D\&I} (S38, S39, S66, S94, S135, S157, S259, S349, I5, I11), commending the survey and the mentorship programs such as, Outreachy and Google Summer of Code (I5, I11).

\textbf{Ways to improve the D\&I initiative}.
Participants provided four ideas for the ASF to address two (out of the eight) negative feedback categories and improve the D\&I initiative. 

To address broader D\&I issues, participants suggested the community to \textit{also consider other diversity aspects besides gender}, like the contributors' sexual orientation (S142, S622), skin color (S142), political affiliation (S142), religion (S142), access to technology (e.g., Internet, phones, computers) (S142), ethnicity (S181, S340, S398, S622, I2), economic factors (S142, S340), and languages spoken (S142, S181, S340). 
In order to put D\&I into practice and have the community take it more seriously, participants claimed the community needs to be more proactive by ``increas[ing] rates of contributor[s] from minority groups", for example ``getting more contributors who are not just males" (I11). Another idea was to \textit{have minority groups as speakers in events} by having specific ``calls for speaker proposals for people who are female or from lower ethnic groups" (I2, I11).
Finally, participants reported that the ASF should \textit{be more inclusive to contributors who are already in}, so ``before trying to get more people, start with the people that are already in the community and help them to feel more included" (S334).

\MyBox{\textbf{D\&I Initiative.} There are mixed feelings, while some appreciate the efforts that the ASF has undertaken to improve D\&I, others are skeptical, feeling that current efforts are focused mainly on gender, is only `lip service', and cause polarization.}
\section{Discussion}
\label{sec:discussion}
\noindent\textbf{Diversity is not only about gender.} Participants in our study perceived that the ASF initiative to improve diversity is mostly focused on gender, missing other struggles of individuals with other diversity aspects such as, language skills, race, ethnicity, age, and other attributes that differentiate people (see Section \ref{sec:feedback}). For example, contributors who were not confident English speakers were more likely to struggle in following technical discussions. This is in line with \citet{steinmacher2015social}'s study, which showed that non-native English speakers face communication barriers. Considering that non-native English speakers can be discriminated against \cite{davidson2014older}, they can lose their self-confidence, feel disengaged, and quit. This raises the importance of taking into account this minority (20\% of our survey respondents are non-confident English speakers) and the language barrier they face which can hinder their participation and sense of belonging \cite{steinmacher2015social}. But proficiency in English is only one attribute that creates  barriers---a recent work compared the acceptance of contributors from countries with different levels of human development~\cite{furtado2020successful}---research needs to also focus on other aspects of diversity. For instance, given the new landscape of OSS~\cite{Steinmacher.Teenager:2017}, projects are now counting on a mix of paid and volunteer contributors working either in standalone OSS projects or open source arms of commercial companies~\cite{maenpaa2018organizing, shah2006motivation}, resulting in a variety of motivations and pathways for contributing~\cite{von2012carrots, gerosa2021motivation, lee2017understanding, trinkenreich2020hidden}. Therefore, compensation type is also a diversity attribute that needs to be studied. The hybrid OSS landscape is different from traditional OSS structure and philosophy as well as different from corporate settings studied by management literature~\cite{peters2003managing,nakatsu2009comparative,georgieva2008best}. Here, we take a first step in this direction. 

\noindent\textbf{Lets talk intersectionality.}
By using more than one lens, our results shed light on potential intersectionality effect among different background attributes. Intersectionality can heighten the perception of being \textsc{Patronized}, the expectation of being a \textsc{Care taker}, impact the perception of being \textsc{Represented} and the difficulty of building a \textsc{Network}. Our results show that those contributors \review{in the gender-minority group} who are not confident in English report feeling patronized by their community (see Table \ref{tab:genderStereotypingOLR}). Other analyses could be done to understand how these factors interact. For example, how about gender minority contributors who are also not confident in English? Is their feeling of being patronized exacerbated by the interplay of stereotypes associated to both gender and English proficiency? Since diversity is a multidimensional construct, a unidimensional view does not paint a complete picture. We need to start looking at diversity from a multidimensional perspective, one that investigates the interplay between the different background attributes \review{\cite{albusays2021diversity}} and the consequences of such interplay on the D\&I initiative.

More recently, research work in computing and collaborative work started investigating the intersectional experience of Black women \cite{rankin2020intersectional, erete2021can}.  Similarly, recent work in HCI demonstrates how intersectionality can be applied to designing technology \cite{rankin2020seat} and used as a model for exposing oppression systems and understanding the inner-works of power within the HCI field \cite{together2021activism, erete2021can, rankin2021resisting}. These works can guide future research on intersectionality in OSS.   

\noindent\textbf{Forming the full picture.}
Gender bias has been studied in different context in OSS showing that contributors who identify as women are subject to stereotypes, biases and social barriers~\cite{imtiaz2019,Terell-2017,kofink2015}. While prior research has found women to be seen as motherly, warm, and nurturing~\cite{mckinnon2020perceptions, carli2016stereotypes, heilman2007women, mitchell2018gender}, our data showed that women (in our dataset) did not feel pigenholed into these stereotypes (Section~\ref{sec:role_streo}). These findings align with prior empirical work by \citet{Lee.Carver:2019}. A possible explanation, as pointed by previous research on women acceptance rate~\cite{Terell-2017}, is survival bias. Given that our respondents are active contributors, most of them with 3+ years in the ASF (73.35\%), our results may reflect the view of those who survive the barriers to contribution and biases. This points to a call for studies that focus on understanding the perception of those individuals from minority groups who left the community and those who do not think that contributing to OSS is feasible for them.

\noindent\textbf{Hand-in-hand with contributors to define the D\&I initiative.} Our results showed that some contributors reported D\&I to be a real problem, while others indicated that D\&I should not receive much attention. Besides, while some contributors support and appreciate the D\&I initiative implemented at the ASF, others have several critiques about the focus, management, and consequences of the initiative. This highlights the importance of making the community aware of the state of D\&I and the impact that the lack of D\&I can cause to people from the many different minority groups. This also highlights the importance of investing in designing and creating an appropriate D\&I initiative that is based on the contributors' needs and refined by getting feedback from different stakeholders. Further, why this initiative is created and how it would be implemented should be communicated back to the community with a process to collect constant feedback. OSS community managers can use the GQM+Strategies approach \cite{basili2014aligning}, and start by understanding the community needs and defining the D\&I goals to be achieved, then cascading those goals into metrics and strategies. The use of metrics can help keep track and control the success or failure of the initiative and associated D\&I goals through a measurement system.
\vspace{-2pt}
\section{Threats to validity}
As any empirical study, this work also has limitations and
threats to validity, which we present in this section. 

\textbf{Construct validity} relates to the constructs used in our study. 
Incorrect questions in our survey or interviews can lead to incorrect measurements. To mitigate this threat, we reused questionnaires where we could \cite{Lee.Carver:2019, 2016ASFSurvey} and collaborated with the ASF D\&I committee to design the instruments. Another threat can arise in the qualitative analysis process. To avoid misinterpretation in the qualitative coding of the data, we used the constant comparison method. As new codes emerged, we compared it with the existing code set and met frequently with the research team to discuss and clarify the codes. The code set generated from the survey results were then compared with the interview data.

\textbf{Internal validity} is related to our ability to capture the reality as close as possible, which in our case is accurately capturing contributors perspective of the state of D\&I and the D\&I initiative. The first threat might arise if we have a biased sampling of the ASF contributors. We believe this limitation was low as we deployed the survey widely receiving 600+ respondents who represented a wide set of demographics. We also leveraged mixed method and survey and interview data to better understand contributors' perspectives. 

\textbf{Conclusion validity}
We focused our study on current ASF contributors. This might result in survival bias, potentially painting a more ``optimistic'' picture. The perceptions of contributors who tried contributing and quit may differ. Another potential threat can be
the imbalance in the number of participants from each background attribute, which might have resulted in fewer results were we found statistically significant differences between groups. Additionally, our results can suffer from self-selection biases \review{and no-response bias} as participation was voluntary \review{we might have missed out on the point of view of contributors who did not choose to respond to the survey}. It is also possible that those who were not proficient in English or those in the minority group knowing that the survey came from the ASF might not have volunteered. 

\textbf{External Validity} 
Even though, the ASF is one of the largest OSS foundations, this study case might not be generalizable to all OSS projects and foundations. We believe however, that our results are representative since the ASF includes 300+ projects across multiple domains and thousands of contributors. The survey was answered by 624 ASF contributors resulting in a response rate of 8.5\% based on a considered total community size of 7500 contributors.  
\vspace{-2pt}
\section{Conclusion}
 \label{sec:conclusion}
 In this work, we investigated contributor perceptions of the state of D\&I and the D\&I initiative at the ASF through 6 different lenses namely gender, education, English proficiency, seniority at ASF, compensation type and country of residence. Our findings show that in addition to gender, others lenses matter in contributors' perception of the state of D\&I (e.g., seniority at ASF, English proficiency, compensation type). 
 Diversity is a multidimensional construct and research needs to use a broader lens to understand the state of D\&I in OSS. Future work needs to also start to investigate how intersectionality plays a role for those who are in the minority.
 
Making substantive changes to a large, decentralized, heterogeneous community like the ASF is not easy. Some felt that D\&I issues are a made-up construct and diversity attributes are invisible in OSS. While others recognized that ``the Foundation operates largely as an `old boys club'(S618).  
While some contributors appreciate the ASF's D\&I effort, some were more critical about the scope of the initiative, the polarization it creates, and the way it was put into practice considering it merely ``to be lip service''(S32).

To truly create an inclusive community and systematize the D\&I efforts, it is important to raise awareness among those who are in `privileged' positions about the impact of biases on individuals who are in the minority in order to make them `allies' in the quest to make the ASF more diverse. It is heartening that the ASF has committed to its D\&I initiative and that the community is rooting for its success: ``It [the D\&I initiative] is in its infancy at the ASF, but I am glad it is happening''(S349).
\begin{acks}
We thank all our survey respondents and our interview participants for their time and insight. This work is partially supported by the National Science Foundation (grants 1901031, 1815486, 1815503 and 1900903) and CNPq (grant \#313067/2020-1).
\end{acks}


\bibliographystyle{ACM-Reference-Format}
\bibliography{biblio}


\begin{thebibliography}{84}


\ifx \showCODEN    \undefined \def \showCODEN     #1{\unskip}     \fi
\ifx \showDOI      \undefined \def \showDOI       #1{#1}\fi
\ifx \showISBNx    \undefined \def \showISBNx     #1{\unskip}     \fi
\ifx \showISBNxiii \undefined \def \showISBNxiii  #1{\unskip}     \fi
\ifx \showISSN     \undefined \def \showISSN      #1{\unskip}     \fi
\ifx \showLCCN     \undefined \def \showLCCN      #1{\unskip}     \fi
\ifx \shownote     \undefined \def \shownote      #1{#1}          \fi
\ifx \showarticletitle \undefined \def \showarticletitle #1{#1}   \fi
\ifx \showURL      \undefined \def \showURL       {\relax}        \fi
\providecommand\bibfield[2]{#2}
\providecommand\bibinfo[2]{#2}
\providecommand\natexlab[1]{#1}
\providecommand\showeprint[2][]{arXiv:#2}

\bibitem[\protect\citeauthoryear{??}{Coc}{[n.d.]}]%
        {Coc}
 \bibinfo{year}{[n.d.]}\natexlab{}.
\newblock \bibinfo{title}{Your Code of Conduct}.
\newblock
  \bibinfo{howpublished}{\url{https://opensource.guide/code-of-conduct/}}.
\newblock
\newblock
\shownote{Accessed: 2021-04-26.}


\bibitem[\protect\citeauthoryear{Albusays, Bjorn, Dabbish, Ford, Murphy-Hill,
  Serebrenik, and Storey}{Albusays et~al\mbox{.}}{2021}]%
        {albusays2021diversity}
\bibfield{author}{\bibinfo{person}{Khaled Albusays}, \bibinfo{person}{Pernille
  Bjorn}, \bibinfo{person}{Laura Dabbish}, \bibinfo{person}{Denae Ford},
  \bibinfo{person}{Emerson Murphy-Hill}, \bibinfo{person}{Alexander
  Serebrenik}, {and} \bibinfo{person}{Margaret-Anne Storey}.}
  \bibinfo{year}{2021}\natexlab{}.
\newblock \showarticletitle{The diversity crisis in software development}.
\newblock \bibinfo{journal}{\emph{IEEE Software}} \bibinfo{volume}{38},
  \bibinfo{number}{2} (\bibinfo{year}{2021}), \bibinfo{pages}{19--25}.
\newblock


\bibitem[\protect\citeauthoryear{Balali, Steinmacher, Annamalai, Sarma, and
  Gerosa}{Balali et~al\mbox{.}}{2018}]%
        {balali2018newcomers}
\bibfield{author}{\bibinfo{person}{Sogol Balali}, \bibinfo{person}{Igor
  Steinmacher}, \bibinfo{person}{Umayal Annamalai}, \bibinfo{person}{Anita
  Sarma}, {and} \bibinfo{person}{Marco~Aurelio Gerosa}.}
  \bibinfo{year}{2018}\natexlab{}.
\newblock \showarticletitle{Newcomers’ barriers... is that all? an analysis
  of mentors’ and newcomers’ barriers in OSS projects}.
\newblock \bibinfo{journal}{\emph{Computer Supported Cooperative Work (CSCW)}}
  \bibinfo{volume}{27}, \bibinfo{number}{3-6} (\bibinfo{year}{2018}),
  \bibinfo{pages}{679--714}.
\newblock


\bibitem[\protect\citeauthoryear{Barreto and Ellemers}{Barreto and
  Ellemers}{2015}]%
        {barreto2015detecting}
\bibfield{author}{\bibinfo{person}{Manuela Barreto} {and}
  \bibinfo{person}{Naomi Ellemers}.} \bibinfo{year}{2015}\natexlab{}.
\newblock \showarticletitle{Detecting and experiencing prejudice: New answers
  to old questions}.
\newblock \bibinfo{journal}{\emph{Advances in experimental social psychology}}
  \bibinfo{volume}{52} (\bibinfo{year}{2015}), \bibinfo{pages}{139--219}.
\newblock


\bibitem[\protect\citeauthoryear{Basili, Trendowicz, Kowalczyk, Heidrich,
  Seaman, Mnch, and Rombach}{Basili et~al\mbox{.}}{2014}]%
        {basili2014aligning}
\bibfield{author}{\bibinfo{person}{Victor Basili}, \bibinfo{person}{Adam
  Trendowicz}, \bibinfo{person}{Martin Kowalczyk}, \bibinfo{person}{Jens
  Heidrich}, \bibinfo{person}{Carolyn Seaman}, \bibinfo{person}{Jrgen Mnch},
  {and} \bibinfo{person}{Dieter Rombach}.} \bibinfo{year}{2014}\natexlab{}.
\newblock \bibinfo{title}{Aligning Organizations Through Measurement: The GQM+
  Strategies Approach}.
\newblock
\newblock


\bibitem[\protect\citeauthoryear{Bosu and Sultana}{Bosu and Sultana}{2019}]%
        {bosu2019diversity}
\bibfield{author}{\bibinfo{person}{Amiangshu Bosu} {and}
  \bibinfo{person}{Kazi~Zakia Sultana}.} \bibinfo{year}{2019}\natexlab{}.
\newblock \showarticletitle{Diversity and inclusion in open source software
  (OSS) projects: Where do we stand?}. In \bibinfo{booktitle}{\emph{2019
  ACM/IEEE International Symposium on Empirical Software Engineering and
  Measurement (ESEM)}}. \bibinfo{publisher}{IEEE}, \bibinfo{pages}{1--11}.
\newblock


\bibitem[\protect\citeauthoryear{Calvo}{Calvo}{2020}]%
        {calvo2020}
\bibfield{author}{\bibinfo{person}{Dafne Calvo}.}
  \bibinfo{year}{2020}\natexlab{}.
\newblock \showarticletitle{The (in) visible barriers to free software:
  Inequalities in online communities in Spain}.
\newblock \bibinfo{journal}{\emph{Studies in Communication Sciences}}
  (\bibinfo{year}{2020}), \bibinfo{pages}{1--16}.
\newblock


\bibitem[\protect\citeauthoryear{Canedo, Bonif{\'a}cio, Okimoto, Serebrenik,
  Pinto, and Monteiro}{Canedo et~al\mbox{.}}{2020}]%
        {canedo2020work}
\bibfield{author}{\bibinfo{person}{Edna~Dias Canedo}, \bibinfo{person}{Rodrigo
  Bonif{\'a}cio}, \bibinfo{person}{M{\'a}rcio~Vinicius Okimoto},
  \bibinfo{person}{Alexander Serebrenik}, \bibinfo{person}{Gustavo Pinto},
  {and} \bibinfo{person}{Eduardo Monteiro}.} \bibinfo{year}{2020}\natexlab{}.
\newblock \showarticletitle{Work Practices and Perceptions from Women Core
  Developers in OSS Communities}. In \bibinfo{booktitle}{\emph{Proceedings of
  the 14th ACM/IEEE International Symposium on Empirical Software Engineering
  and Measurement (ESEM)}}. \bibinfo{pages}{1--11}.
\newblock


\bibitem[\protect\citeauthoryear{Carli, Alawa, Lee, Zhao, and Kim}{Carli
  et~al\mbox{.}}{2016}]%
        {carli2016stereotypes}
\bibfield{author}{\bibinfo{person}{Linda~L Carli}, \bibinfo{person}{Laila
  Alawa}, \bibinfo{person}{YoonAh Lee}, \bibinfo{person}{Bei Zhao}, {and}
  \bibinfo{person}{Elaine Kim}.} \bibinfo{year}{2016}\natexlab{}.
\newblock \showarticletitle{Stereotypes about gender and science: Women$\ne$
  scientists}.
\newblock \bibinfo{journal}{\emph{Psychology of Women Quarterly}}
  \bibinfo{volume}{40}, \bibinfo{number}{2} (\bibinfo{year}{2016}),
  \bibinfo{pages}{244--260}.
\newblock


\bibitem[\protect\citeauthoryear{Catolino, Palomba, Tamburri, Serebrenik, and
  Ferrucci}{Catolino et~al\mbox{.}}{2019}]%
        {catolino2019gender}
\bibfield{author}{\bibinfo{person}{Gemma Catolino}, \bibinfo{person}{Fabio
  Palomba}, \bibinfo{person}{Damian~A Tamburri}, \bibinfo{person}{Alexander
  Serebrenik}, {and} \bibinfo{person}{Filomena Ferrucci}.}
  \bibinfo{year}{2019}\natexlab{}.
\newblock \showarticletitle{Gender diversity and community smells: insights
  from the trenches}.
\newblock \bibinfo{journal}{\emph{IEEE Software}} \bibinfo{volume}{37},
  \bibinfo{number}{1} (\bibinfo{year}{2019}), \bibinfo{pages}{10--16}.
\newblock


\bibitem[\protect\citeauthoryear{Cuddy, Fiske, and Glick}{Cuddy
  et~al\mbox{.}}{2008}]%
        {cuddy2008warmth}
\bibfield{author}{\bibinfo{person}{Amy~JC Cuddy}, \bibinfo{person}{Susan~T
  Fiske}, {and} \bibinfo{person}{Peter Glick}.}
  \bibinfo{year}{2008}\natexlab{}.
\newblock \showarticletitle{Warmth and competence as universal dimensions of
  social perception: The stereotype content model and the BIAS map}.
\newblock \bibinfo{journal}{\emph{Advances in experimental social psychology}}
  \bibinfo{volume}{40} (\bibinfo{year}{2008}), \bibinfo{pages}{61--149}.
\newblock


\bibitem[\protect\citeauthoryear{David and Shapiro}{David and Shapiro}{2008}]%
        {David2008IEP}
\bibfield{author}{\bibinfo{person}{Paul~A. David} {and}
  \bibinfo{person}{Joseph~S. Shapiro}.} \bibinfo{year}{2008}\natexlab{}.
\newblock \showarticletitle{Community-based production of open-source software:
  What do we know about the developers who participate?}
\newblock \bibinfo{journal}{\emph{Information Economics and Policy}}
  \bibinfo{volume}{20}, \bibinfo{number}{4} (\bibinfo{year}{2008}),
  \bibinfo{pages}{364--398}.
\newblock
\showISSN{0167-6245}
\urldef\tempurl%
\url{https://doi.org/10.1016/j.infoecopol.2008.10.001}
\showDOI{\tempurl}


\bibitem[\protect\citeauthoryear{Davidson, Naik, Mannan, Azarbakht, and
  Jensen}{Davidson et~al\mbox{.}}{2014}]%
        {davidson2014older}
\bibfield{author}{\bibinfo{person}{Jennifer~L Davidson},
  \bibinfo{person}{Rithika Naik}, \bibinfo{person}{Umme~Ayda Mannan},
  \bibinfo{person}{Amir Azarbakht}, {and} \bibinfo{person}{Carlos Jensen}.}
  \bibinfo{year}{2014}\natexlab{}.
\newblock \showarticletitle{On older adults in free/open source software:
  reflections of contributors and community leaders}. In
  \bibinfo{booktitle}{\emph{2014 IEEE Symposium on Visual Languages and
  Human-Centric Computing (VL/HCC)}}. IEEE, \bibinfo{pages}{93--100}.
\newblock


\bibitem[\protect\citeauthoryear{demographics}{demographics}{2021}]%
        {opendemographicsdoc}
\bibfield{author}{\bibinfo{person}{Open demographics}.}
  \bibinfo{year}{2021}\natexlab{}.
\newblock \bibinfo{title}{{Open demographics documentation}}.
\newblock
  \bibinfo{howpublished}{\url{http://nikkistevens.com/open-demographics/}}.
\newblock
\newblock
\shownote{[Online; accessed 2020-12-30].}


\bibitem[\protect\citeauthoryear{Denmark and Paludi}{Denmark and
  Paludi}{2007}]%
        {denmark2007psychology}
\bibfield{author}{\bibinfo{person}{Florence Denmark} {and}
  \bibinfo{person}{Michele~Antoinette Paludi}.}
  \bibinfo{year}{2007}\natexlab{}.
\newblock \bibinfo{booktitle}{\emph{Psychology of women: Handbook of issues and
  theories}}.
\newblock \bibinfo{publisher}{Greenwood Publishing Group}.
\newblock


\bibitem[\protect\citeauthoryear{Erete, Rankin, and Thomas}{Erete
  et~al\mbox{.}}{2021}]%
        {erete2021can}
\bibfield{author}{\bibinfo{person}{Sheena Erete}, \bibinfo{person}{Yolanda~A
  Rankin}, {and} \bibinfo{person}{Jakita~O Thomas}.}
  \bibinfo{year}{2021}\natexlab{}.
\newblock \showarticletitle{I Can't Breathe: Reflections from Black Women in
  CSCW and HCI}.
\newblock \bibinfo{journal}{\emph{Proceedings of the ACM on Human-Computer
  Interaction}} \bibinfo{volume}{4}, \bibinfo{number}{CSCW3}
  (\bibinfo{year}{2021}), \bibinfo{pages}{1--23}.
\newblock


\bibitem[\protect\citeauthoryear{Feller and Fitzgerald}{Feller and
  Fitzgerald}{2000}]%
        {feller2000framework}
\bibfield{author}{\bibinfo{person}{Joseph Feller} {and} \bibinfo{person}{Brian
  Fitzgerald}.} \bibinfo{year}{2000}\natexlab{}.
\newblock \showarticletitle{A framework analysis of the open source software
  development paradigm}. In \bibinfo{booktitle}{\emph{ICIS 2000 proceedings of
  the twenty first international conference on information systems}}.
  Association for Information Systems (AIS), \bibinfo{pages}{58--69}.
\newblock


\bibitem[\protect\citeauthoryear{Fiske}{Fiske}{2015}]%
        {fiske2015intergroup}
\bibfield{author}{\bibinfo{person}{Susan~T Fiske}.}
  \bibinfo{year}{2015}\natexlab{}.
\newblock \showarticletitle{Intergroup biases: A focus on stereotype content}.
\newblock \bibinfo{journal}{\emph{Current opinion in behavioral sciences}}
  \bibinfo{volume}{3} (\bibinfo{year}{2015}), \bibinfo{pages}{45--50}.
\newblock


\bibitem[\protect\citeauthoryear{Fiske, Cuddy, and Glick}{Fiske
  et~al\mbox{.}}{2007}]%
        {fiske2007universal}
\bibfield{author}{\bibinfo{person}{Susan~T Fiske}, \bibinfo{person}{Amy~JC
  Cuddy}, {and} \bibinfo{person}{Peter Glick}.}
  \bibinfo{year}{2007}\natexlab{}.
\newblock \showarticletitle{Universal dimensions of social cognition: Warmth
  and competence}.
\newblock \bibinfo{journal}{\emph{Trends in cognitive sciences}}
  \bibinfo{volume}{11}, \bibinfo{number}{2} (\bibinfo{year}{2007}),
  \bibinfo{pages}{77--83}.
\newblock


\bibitem[\protect\citeauthoryear{Foundation}{Foundation}{1999}]%
        {ASFwebsite}
\bibfield{author}{\bibinfo{person}{Apache~Software Foundation}.}
  \bibinfo{year}{1999}\natexlab{}.
\newblock \bibinfo{title}{{Apache Software Foundation}}.
\newblock \bibinfo{howpublished}{\url{https://www.apache.org/}}.
\newblock
\newblock
\shownote{[Online; accessed 2019-05-17].}


\bibitem[\protect\citeauthoryear{Foundation}{Foundation}{2016}]%
        {2016ASFSurvey}
\bibfield{author}{\bibinfo{person}{Apache~Software Foundation}.}
  \bibinfo{year}{2016}\natexlab{}.
\newblock \bibinfo{title}{{ASF Committer Diversity Survey}}.
\newblock
  \bibinfo{howpublished}{\url{https://cwiki.apache.org/confluence/display/COMDEV/ASF+Committer+Diversity+Survey+-+2016}}.
\newblock
\newblock
\shownote{[Online; accessed 2020-12-30].}


\bibitem[\protect\citeauthoryear{Foundation}{Foundation}{2019}]%
        {EDIgroup}
\bibfield{author}{\bibinfo{person}{Apache~Software Foundation}.}
  \bibinfo{year}{2019}\natexlab{}.
\newblock \bibinfo{title}{{Promoting and Studying Diversity and Inclusion in
  Open Source}}.
\newblock \bibinfo{howpublished}{\url{http://diversity.apache.org/}}.
\newblock
\newblock
\shownote{[Online; accessed 2019-06-10].}


\bibitem[\protect\citeauthoryear{Foundation}{Foundation}{2020}]%
        {linuxFoundationDAndI}
\bibfield{author}{\bibinfo{person}{The~Linux Foundation}.}
  \bibinfo{year}{2020}\natexlab{}.
\newblock \bibinfo{title}{Addressing Diversity and Inclusivity}.
\newblock
\newblock
\urldef\tempurl%
\url{https://www.linuxfoundation.org/en/about/diversity-inclusivity/}
\showURL{%
\tempurl}
\newblock
\shownote{[Online; accessed 29-March-2021].}


\bibitem[\protect\citeauthoryear{Furtado, Cartaxo, Treude, and Pinto}{Furtado
  et~al\mbox{.}}{2020}]%
        {furtado2020successful}
\bibfield{author}{\bibinfo{person}{Leonardo Furtado}, \bibinfo{person}{Bruno
  Cartaxo}, \bibinfo{person}{Christoph Treude}, {and} \bibinfo{person}{Gustavo
  Pinto}.} \bibinfo{year}{2020}\natexlab{}.
\newblock \showarticletitle{How Successful Are Open Source Contributions From
  Countries with Different Levels of Human Development?}
\newblock \bibinfo{journal}{\emph{IEEE Software}} (\bibinfo{year}{2020}).
\newblock


\bibitem[\protect\citeauthoryear{Georgieva and Allan}{Georgieva and
  Allan}{2008}]%
        {georgieva2008best}
\bibfield{author}{\bibinfo{person}{Svetla Georgieva} {and}
  \bibinfo{person}{George Allan}.} \bibinfo{year}{2008}\natexlab{}.
\newblock \showarticletitle{Best Practices in Project Management Through a
  Grounded Theory Lens.}
\newblock \bibinfo{journal}{\emph{Electronic Journal of Business Research
  Methods}} \bibinfo{volume}{6}, \bibinfo{number}{1} (\bibinfo{year}{2008}).
\newblock


\bibitem[\protect\citeauthoryear{Gerosa, Wiese, Trinkenreich, Link, Robles,
  Treude, Steinmacher, and Sarma}{Gerosa et~al\mbox{.}}{2021}]%
        {gerosa2021motivation}
\bibfield{author}{\bibinfo{person}{Marco Gerosa}, \bibinfo{person}{Igor Wiese},
  \bibinfo{person}{Bianca Trinkenreich}, \bibinfo{person}{Georg Link},
  \bibinfo{person}{Gregorio Robles}, \bibinfo{person}{Christoph Treude},
  \bibinfo{person}{Igor Steinmacher}, {and} \bibinfo{person}{Anita Sarma}.}
  \bibinfo{year}{2021}\natexlab{}.
\newblock \showarticletitle{The Shifting Sands of Motivation: Revisiting What
  Drives Contributors in Open Source}. In \bibinfo{booktitle}{\emph{Proceedings
  of the 43rd International Conference on Software Engineering (ICSE)}}.
\newblock


\bibitem[\protect\citeauthoryear{Ghosh, Glott, Krieger, and Robles}{Ghosh
  et~al\mbox{.}}{2002}]%
        {Ghosh2002IIIM}
\bibfield{author}{\bibinfo{person}{R. Ghosh}, \bibinfo{person}{A. Glott},
  \bibinfo{person}{B. Krieger}, {and} \bibinfo{person}{B. Robles}.}
  \bibinfo{year}{2002}\natexlab{}.
\newblock \bibinfo{title}{Free/Libre and Open Source Software: Survey and Study
  ({FLOSS}), Final Report, Part IV: Survey of Developers}.
\newblock
\newblock


\bibitem[\protect\citeauthoryear{Gila, Jaafa, Omar, and Tunio}{Gila
  et~al\mbox{.}}{2014}]%
        {gila2014impact}
\bibfield{author}{\bibinfo{person}{Abdul~Rehman Gila},
  \bibinfo{person}{Jafreezal Jaafa}, \bibinfo{person}{Mazni Omar}, {and}
  \bibinfo{person}{Muhammad~Zahid Tunio}.} \bibinfo{year}{2014}\natexlab{}.
\newblock \showarticletitle{Impact of personality and gender diversity on
  software development teams' performance}. In \bibinfo{booktitle}{\emph{2014
  International Conference on Computer, Communications, and Control Technology
  (I4CT)}}. IEEE, \bibinfo{pages}{261--265}.
\newblock


\bibitem[\protect\citeauthoryear{Guizani, Trinkenreich, Castro-Guzman,
  Steinmacher, Gerosa, and Sarma}{Guizani et~al\mbox{.}}{2022}]%
        {suppdoc}
\bibfield{author}{\bibinfo{person}{Mariam Guizani}, \bibinfo{person}{Bianca
  Trinkenreich}, \bibinfo{person}{Aileen~Abril Castro-Guzman},
  \bibinfo{person}{Igor Steinmacher}, \bibinfo{person}{Marco~A. Gerosa}, {and}
  \bibinfo{person}{Anita Sarma}.} \bibinfo{year}{Feb, 2022}\natexlab{}.
\newblock \bibinfo{title}{{Supplemental Document for Perceptions of the State
  of D\&I and D\&I Initiative in the ASF}}.
\newblock \bibinfo{howpublished}{Available at
  \url{https://figshare.com/s/888e88758dee6b9a9475}}.
\newblock


\bibitem[\protect\citeauthoryear{Hagerty, Williams, Coyne, and Early}{Hagerty
  et~al\mbox{.}}{1996}]%
        {hagerty1996sense}
\bibfield{author}{\bibinfo{person}{Bonnie~M Hagerty}, \bibinfo{person}{Reg~A
  Williams}, \bibinfo{person}{James~C Coyne}, {and} \bibinfo{person}{Margaret~R
  Early}.} \bibinfo{year}{1996}\natexlab{}.
\newblock \showarticletitle{Sense of belonging and indicators of social and
  psychological functioning}.
\newblock \bibinfo{journal}{\emph{Archives of psychiatric nursing}}
  \bibinfo{volume}{10}, \bibinfo{number}{4} (\bibinfo{year}{1996}),
  \bibinfo{pages}{235--244}.
\newblock


\bibitem[\protect\citeauthoryear{Harrell}{Harrell}{2015}]%
        {harrell2015ordinal}
\bibfield{author}{\bibinfo{person}{Frank~E Harrell}.}
  \bibinfo{year}{2015}\natexlab{}.
\newblock \showarticletitle{Ordinal logistic regression}.
\newblock In \bibinfo{booktitle}{\emph{Regression modeling strategies}}.
  \bibinfo{publisher}{Springer}, \bibinfo{pages}{311--325}.
\newblock


\bibitem[\protect\citeauthoryear{Heilman and Okimoto}{Heilman and
  Okimoto}{2007}]%
        {heilman2007women}
\bibfield{author}{\bibinfo{person}{Madeline~E Heilman} {and}
  \bibinfo{person}{Tyler~G Okimoto}.} \bibinfo{year}{2007}\natexlab{}.
\newblock \showarticletitle{Why are women penalized for success at male tasks?:
  the implied communality deficit.}
\newblock \bibinfo{journal}{\emph{Journal of applied psychology}}
  \bibinfo{volume}{92}, \bibinfo{number}{1} (\bibinfo{year}{2007}),
  \bibinfo{pages}{81}.
\newblock


\bibitem[\protect\citeauthoryear{Hoyle and Crawford}{Hoyle and
  Crawford}{1994}]%
        {hoyle1994use}
\bibfield{author}{\bibinfo{person}{Rick~H Hoyle} {and} \bibinfo{person}{Anne~M
  Crawford}.} \bibinfo{year}{1994}\natexlab{}.
\newblock \showarticletitle{Use of individual-level data to investigate group
  phenomena issues and strategies}.
\newblock \bibinfo{journal}{\emph{Small Group Research}} \bibinfo{volume}{25},
  \bibinfo{number}{4} (\bibinfo{year}{1994}), \bibinfo{pages}{464--485}.
\newblock


\bibitem[\protect\citeauthoryear{Imtiaz, Middleton, Chakraborty, Robson, Bai,
  and Murphy-Hill}{Imtiaz et~al\mbox{.}}{2019}]%
        {imtiaz2019}
\bibfield{author}{\bibinfo{person}{Nasif Imtiaz}, \bibinfo{person}{Justin
  Middleton}, \bibinfo{person}{Joymallya Chakraborty}, \bibinfo{person}{Neill
  Robson}, \bibinfo{person}{Gina Bai}, {and} \bibinfo{person}{Emerson
  Murphy-Hill}.} \bibinfo{year}{2019}\natexlab{}.
\newblock \showarticletitle{Investigating the effects of gender bias on
  GitHub}. In \bibinfo{booktitle}{\emph{2019 IEEE/ACM 41st International
  Conference on Software Engineering (ICSE)}}. IEEE, \bibinfo{pages}{700--711}.
\newblock


\bibitem[\protect\citeauthoryear{Izquierdo, Huesman, Serebrenik, and
  Robles}{Izquierdo et~al\mbox{.}}{2018}]%
        {izquierdo2018openstack}
\bibfield{author}{\bibinfo{person}{Daniel Izquierdo}, \bibinfo{person}{Nicole
  Huesman}, \bibinfo{person}{Alexander Serebrenik}, {and}
  \bibinfo{person}{Gregorio Robles}.} \bibinfo{year}{2018}\natexlab{}.
\newblock \showarticletitle{Openstack gender diversity report}.
\newblock \bibinfo{journal}{\emph{IEEE Software}} \bibinfo{volume}{36},
  \bibinfo{number}{1} (\bibinfo{year}{2018}), \bibinfo{pages}{28--33}.
\newblock


\bibitem[\protect\citeauthoryear{Kaplan}{Kaplan}{1994}]%
        {kaplan1994woman}
\bibfield{author}{\bibinfo{person}{Laura~Duhan Kaplan}.}
  \bibinfo{year}{1994}\natexlab{}.
\newblock \showarticletitle{Woman as caretaker: An archetype that supports
  patriarchal militarism}.
\newblock \bibinfo{journal}{\emph{Hypatia}} \bibinfo{volume}{9},
  \bibinfo{number}{2} (\bibinfo{year}{1994}), \bibinfo{pages}{123--133}.
\newblock


\bibitem[\protect\citeauthoryear{Kofink}{Kofink}{2015}]%
        {kofink2015}
\bibfield{author}{\bibinfo{person}{Andrew Kofink}.}
  \bibinfo{year}{2015}\natexlab{}.
\newblock \showarticletitle{Contributions of the under-appreciated: gender bias
  in an open-source ecology}. In \bibinfo{booktitle}{\emph{Companion
  Proceedings of the 2015 ACM SIGPLAN International Conference on Systems,
  Programming, Languages and Applications: Software for Humanity}}.
  \bibinfo{pages}{83--84}.
\newblock


\bibitem[\protect\citeauthoryear{Lee and Carver}{Lee and Carver}{2019a}]%
        {Lee.Carver:2019}
\bibfield{author}{\bibinfo{person}{Amanda Lee} {and} \bibinfo{person}{Jeffrey
  Carver}.} \bibinfo{year}{2019}\natexlab{a}.
\newblock \showarticletitle{FLOSS Participants' Perceptions about Gender and
  Inclusiveness: A Survey}. In \bibinfo{booktitle}{\emph{41st International
  Conference on Software Engineering}}.
\newblock


\bibitem[\protect\citeauthoryear{Lee and Carver}{Lee and Carver}{2019b}]%
        {lee2019floss}
\bibfield{author}{\bibinfo{person}{Amanda Lee} {and} \bibinfo{person}{Jeffrey~C
  Carver}.} \bibinfo{year}{2019}\natexlab{b}.
\newblock \showarticletitle{FLOSS participants' perceptions about gender and
  inclusiveness: a survey}. In \bibinfo{booktitle}{\emph{2019 IEEE/ACM 41st
  International Conference on Software Engineering (ICSE)}}. IEEE,
  \bibinfo{pages}{677--687}.
\newblock


\bibitem[\protect\citeauthoryear{Lee, Carver, and Bosu}{Lee
  et~al\mbox{.}}{2017}]%
        {lee2017understanding}
\bibfield{author}{\bibinfo{person}{Amanda Lee}, \bibinfo{person}{Jeffrey~C
  Carver}, {and} \bibinfo{person}{Amiangshu Bosu}.}
  \bibinfo{year}{2017}\natexlab{}.
\newblock \showarticletitle{Understanding the impressions, motivations, and
  barriers of one time code contributors to FLOSS projects: a survey}. In
  \bibinfo{booktitle}{\emph{2017 IEEE/ACM 39th International Conference on
  Software Engineering (ICSE)}}. IEEE, \bibinfo{pages}{187--197}.
\newblock


\bibitem[\protect\citeauthoryear{Li, Pandurangan, Frluckaj, and Dabbish}{Li
  et~al\mbox{.}}{2021}]%
        {li2021code}
\bibfield{author}{\bibinfo{person}{Renee Li}, \bibinfo{person}{Pavitthra
  Pandurangan}, \bibinfo{person}{Hana Frluckaj}, {and} \bibinfo{person}{Laura
  Dabbish}.} \bibinfo{year}{2021}\natexlab{}.
\newblock \showarticletitle{Code of Conduct Conversations in Open Source
  Software Projects on Github}.
\newblock \bibinfo{journal}{\emph{Proceedings of the ACM on Human-Computer
  Interaction}} \bibinfo{volume}{5}, \bibinfo{number}{CSCW1}
  (\bibinfo{year}{2021}), \bibinfo{pages}{1--31}.
\newblock


\bibitem[\protect\citeauthoryear{Lim}{Lim}{2008}]%
        {lim2008job}
\bibfield{author}{\bibinfo{person}{Sook Lim}.} \bibinfo{year}{2008}\natexlab{}.
\newblock \showarticletitle{Job satisfaction of information technology workers
  in academic libraries}.
\newblock \bibinfo{journal}{\emph{Library \& Information Science Research}}
  \bibinfo{volume}{30}, \bibinfo{number}{2} (\bibinfo{year}{2008}),
  \bibinfo{pages}{115--121}.
\newblock


\bibitem[\protect\citeauthoryear{Lin and Serebrenik}{Lin and
  Serebrenik}{2016}]%
        {lin2016recognizing}
\bibfield{author}{\bibinfo{person}{Bin Lin} {and} \bibinfo{person}{Alexander
  Serebrenik}.} \bibinfo{year}{2016}\natexlab{}.
\newblock \showarticletitle{Recognizing gender of stack overflow users}. In
  \bibinfo{booktitle}{\emph{Proceedings of the 13th International Conference on
  Mining Software Repositories}}. \bibinfo{pages}{425--429}.
\newblock


\bibitem[\protect\citeauthoryear{M{\"a}enp{\"a}{\"a}, M{\"a}kinen, Kilamo,
  Mikkonen, M{\"a}nnist{\"o}, and Ritala}{M{\"a}enp{\"a}{\"a}
  et~al\mbox{.}}{2018}]%
        {maenpaa2018organizing}
\bibfield{author}{\bibinfo{person}{Hanna M{\"a}enp{\"a}{\"a}},
  \bibinfo{person}{Simo M{\"a}kinen}, \bibinfo{person}{Terhi Kilamo},
  \bibinfo{person}{Tommi Mikkonen}, \bibinfo{person}{Tomi M{\"a}nnist{\"o}},
  {and} \bibinfo{person}{Paavo Ritala}.} \bibinfo{year}{2018}\natexlab{}.
\newblock \showarticletitle{Organizing for openness: six models for developer
  involvement in hybrid OSS projects}.
\newblock \bibinfo{journal}{\emph{Journal of Internet Services and
  Applications}} \bibinfo{volume}{9}, \bibinfo{number}{1}
  (\bibinfo{year}{2018}), \bibinfo{pages}{17}.
\newblock


\bibitem[\protect\citeauthoryear{Marlow, Dabbish, and Herbsleb}{Marlow
  et~al\mbox{.}}{2013}]%
        {Marlow2013CSCW}
\bibfield{author}{\bibinfo{person}{Jennifer Marlow}, \bibinfo{person}{Laura
  Dabbish}, {and} \bibinfo{person}{Jim Herbsleb}.}
  \bibinfo{year}{2013}\natexlab{}.
\newblock \showarticletitle{Impression Formation in Online Peer Production:
  Activity Traces and Personal Profiles in Github}. In
  \bibinfo{booktitle}{\emph{Proceedings of the 2013 Conference on Computer
  Supported Cooperative Work}} \emph{(\bibinfo{series}{CSCW '13})}.
  \bibinfo{publisher}{ACM}, \bibinfo{address}{New York, NY, USA},
  \bibinfo{pages}{117--128}.
\newblock


\bibitem[\protect\citeauthoryear{McKinnon and O’Connell}{McKinnon and
  O’Connell}{2020}]%
        {mckinnon2020perceptions}
\bibfield{author}{\bibinfo{person}{Merryn McKinnon} {and}
  \bibinfo{person}{Christine O’Connell}.} \bibinfo{year}{2020}\natexlab{}.
\newblock \showarticletitle{Perceptions of stereotypes applied to women who
  publicly communicate their STEM work}.
\newblock \bibinfo{journal}{\emph{Humanities and Social Sciences
  Communications}} \bibinfo{volume}{7}, \bibinfo{number}{1}
  (\bibinfo{year}{2020}), \bibinfo{pages}{1--8}.
\newblock


\bibitem[\protect\citeauthoryear{Mendez, Padala, Steine-Hanson, Hilderbrand,
  Horvath, Hill, Simpson, Patil, Sarma, and Burnett}{Mendez
  et~al\mbox{.}}{2018}]%
        {mendez2018open}
\bibfield{author}{\bibinfo{person}{Christopher Mendez},
  \bibinfo{person}{Hema~Susmita Padala}, \bibinfo{person}{Zoe Steine-Hanson},
  \bibinfo{person}{Claudia Hilderbrand}, \bibinfo{person}{Amber Horvath},
  \bibinfo{person}{Charles Hill}, \bibinfo{person}{Logan Simpson},
  \bibinfo{person}{Nupoor Patil}, \bibinfo{person}{Anita Sarma}, {and}
  \bibinfo{person}{Margaret Burnett}.} \bibinfo{year}{2018}\natexlab{}.
\newblock \showarticletitle{Open source barriers to entry, revisited: A
  sociotechnical perspective}. In \bibinfo{booktitle}{\emph{Proceedings of the
  40th International Conference on Software Engineering}}.
  \bibinfo{pages}{1004--1015}.
\newblock


\bibitem[\protect\citeauthoryear{Mitchell and Martin}{Mitchell and
  Martin}{2018}]%
        {mitchell2018gender}
\bibfield{author}{\bibinfo{person}{Kristina~MW Mitchell} {and}
  \bibinfo{person}{Jonathan Martin}.} \bibinfo{year}{2018}\natexlab{}.
\newblock \showarticletitle{Gender bias in student evaluations}.
\newblock \bibinfo{journal}{\emph{PS: Political Science \& Politics}}
  \bibinfo{volume}{51}, \bibinfo{number}{3} (\bibinfo{year}{2018}),
  \bibinfo{pages}{648--652}.
\newblock


\bibitem[\protect\citeauthoryear{Morrison, Pandita, Murphy-Hill, and
  McLaughlin}{Morrison et~al\mbox{.}}{2016}]%
        {morrison2016veteran}
\bibfield{author}{\bibinfo{person}{Patrick Morrison}, \bibinfo{person}{Rahul
  Pandita}, \bibinfo{person}{Emerson Murphy-Hill}, {and} \bibinfo{person}{Anne
  McLaughlin}.} \bibinfo{year}{2016}\natexlab{}.
\newblock \showarticletitle{Veteran developers' contributions and motivations:
  An open source perspective}. In \bibinfo{booktitle}{\emph{2016 IEEE Symposium
  on Visual Languages and Human-Centric Computing (VL/HCC)}}. IEEE,
  \bibinfo{pages}{171--179}.
\newblock


\bibitem[\protect\citeauthoryear{Murakami, Tsunoda, and Uwano}{Murakami
  et~al\mbox{.}}{2017}]%
        {murakami2017wap}
\bibfield{author}{\bibinfo{person}{Yukasa Murakami}, \bibinfo{person}{Masateru
  Tsunoda}, {and} \bibinfo{person}{Hidetake Uwano}.}
  \bibinfo{year}{2017}\natexlab{}.
\newblock \showarticletitle{WAP: Does Reviewer Age Affect Code Review
  Performance?}. In \bibinfo{booktitle}{\emph{2017 IEEE 28th International
  Symposium on Software Reliability Engineering (ISSRE)}}. IEEE,
  \bibinfo{pages}{164--169}.
\newblock


\bibitem[\protect\citeauthoryear{Nafus}{Nafus}{2012}]%
        {nafus2012patches}
\bibfield{author}{\bibinfo{person}{Dawn Nafus}.}
  \bibinfo{year}{2012}\natexlab{}.
\newblock \showarticletitle{‘Patches don’t have gender’: What is not open
  in open source software}.
\newblock \bibinfo{journal}{\emph{New Media \& Society}} \bibinfo{volume}{14},
  \bibinfo{number}{4} (\bibinfo{year}{2012}), \bibinfo{pages}{669--683}.
\newblock


\bibitem[\protect\citeauthoryear{Nakatsu and Iacovou}{Nakatsu and
  Iacovou}{2009}]%
        {nakatsu2009comparative}
\bibfield{author}{\bibinfo{person}{Robbie~T Nakatsu} {and}
  \bibinfo{person}{Charalambos~L Iacovou}.} \bibinfo{year}{2009}\natexlab{}.
\newblock \showarticletitle{A comparative study of important risk factors
  involved in offshore and domestic outsourcing of software development
  projects: A two-panel Delphi study}.
\newblock \bibinfo{journal}{\emph{Information \& Management}}
  \bibinfo{volume}{46}, \bibinfo{number}{1} (\bibinfo{year}{2009}),
  \bibinfo{pages}{57--68}.
\newblock


\bibitem[\protect\citeauthoryear{Ortu, Destefanis, Counsell, Swift, Tonelli,
  and Marchesi}{Ortu et~al\mbox{.}}{2017}]%
        {ortu2017diverse}
\bibfield{author}{\bibinfo{person}{Marco Ortu}, \bibinfo{person}{Giuseppe
  Destefanis}, \bibinfo{person}{Steve Counsell}, \bibinfo{person}{Stephen
  Swift}, \bibinfo{person}{Roberto Tonelli}, {and} \bibinfo{person}{Michele
  Marchesi}.} \bibinfo{year}{2017}\natexlab{}.
\newblock \showarticletitle{How diverse is your team? Investigating gender and
  nationality diversity in GitHub teams}.
\newblock \bibinfo{journal}{\emph{Journal of Software Engineering Research and
  Development}} \bibinfo{volume}{5}, \bibinfo{number}{1}
  (\bibinfo{year}{2017}), \bibinfo{pages}{1--18}.
\newblock


\bibitem[\protect\citeauthoryear{Padala, Mendez, Dias, Steinmacher, Hanson,
  Hilderbrand, Horvath, Hill, Simpson, Burnett, et~al\mbox{.}}{Padala
  et~al\mbox{.}}{2020}]%
        {padala2020gender}
\bibfield{author}{\bibinfo{person}{Susmita~Hema Padala},
  \bibinfo{person}{Christopher~John Mendez}, \bibinfo{person}{Luiz~Felipe
  Dias}, \bibinfo{person}{Igor Steinmacher}, \bibinfo{person}{Zoe~Steine
  Hanson}, \bibinfo{person}{Claudia Hilderbrand}, \bibinfo{person}{Amber
  Horvath}, \bibinfo{person}{Charles Hill}, \bibinfo{person}{Logan~Dale
  Simpson}, \bibinfo{person}{Margaret Burnett}, {et~al\mbox{.}}}
  \bibinfo{year}{2020}\natexlab{}.
\newblock \showarticletitle{How Gender-biased Tools Shape Newcomer Experiences
  in OSS Projects}.
\newblock \bibinfo{journal}{\emph{IEEE Transactions on Software Engineering}}
  (\bibinfo{year}{2020}).
\newblock


\bibitem[\protect\citeauthoryear{Peters}{Peters}{2003}]%
        {peters2003managing}
\bibfield{author}{\bibinfo{person}{Lawrence Peters}.}
  \bibinfo{year}{2003}\natexlab{}.
\newblock \showarticletitle{Managing software professionals}. In
  \bibinfo{booktitle}{\emph{IEMC'03 Proceedings. Managing Technologically
  Driven Organizations: The Human Side of Innovation and Change}}. IEEE,
  \bibinfo{pages}{61--66}.
\newblock


\bibitem[\protect\citeauthoryear{Powell, Hunsinger, and Medlin}{Powell
  et~al\mbox{.}}{2010}]%
        {powell2010}
\bibfield{author}{\bibinfo{person}{Whitney~E Powell}, \bibinfo{person}{D~Scott
  Hunsinger}, {and} \bibinfo{person}{B~Dawn Medlin}.}
  \bibinfo{year}{2010}\natexlab{}.
\newblock \showarticletitle{Gender differences within the open source
  community: An exploratory study}.
\newblock \bibinfo{journal}{\emph{Journal of Information Technology}}
  \bibinfo{volume}{21}, \bibinfo{number}{4} (\bibinfo{year}{2010}),
  \bibinfo{pages}{29--37}.
\newblock


\bibitem[\protect\citeauthoryear{Prana, Ford, Rastogi, Lo, Purandare, and
  Nagappan}{Prana et~al\mbox{.}}{2020}]%
        {prana2020including}
\bibfield{author}{\bibinfo{person}{Gede Artha~Azriadi Prana},
  \bibinfo{person}{Denae Ford}, \bibinfo{person}{Ayushi Rastogi},
  \bibinfo{person}{David Lo}, \bibinfo{person}{Rahul Purandare}, {and}
  \bibinfo{person}{Nachiappan Nagappan}.} \bibinfo{year}{2020}\natexlab{}.
\newblock \showarticletitle{Including Everyone, Everywhere: Understanding
  Opportunities and Challenges of Geographic Gender-Inclusion in OSS}.
\newblock \bibinfo{journal}{\emph{ACM Transactions on Software Engineering and
  Methodology (TOSEM)}} (\bibinfo{year}{2020}).
\newblock


\bibitem[\protect\citeauthoryear{Project}{Project}{2017}]%
        {chaossmetrics}
\bibfield{author}{\bibinfo{person}{CHAOSS Project}.}
  \bibinfo{year}{2017}\natexlab{}.
\newblock \bibinfo{title}{{Diversity and inclusion metrics}}.
\newblock
  \bibinfo{howpublished}{\url{https://github.com/chaoss/wg-diversity-inclusion/tree/master/focus-areas}}.
\newblock
\newblock
\shownote{[Online; accessed 2020-12-30].}


\bibitem[\protect\citeauthoryear{Rankin and Henderson}{Rankin and
  Henderson}{2021}]%
        {rankin2021resisting}
\bibfield{author}{\bibinfo{person}{Yolanda~A Rankin} {and}
  \bibinfo{person}{Kallayah~K Henderson}.} \bibinfo{year}{2021}\natexlab{}.
\newblock \showarticletitle{Resisting Racism in Tech Design: Centering the
  Experiences of Black Youth}.
\newblock \bibinfo{journal}{\emph{Proceedings of the ACM on Human-Computer
  Interaction}} \bibinfo{volume}{5}, \bibinfo{number}{CSCW1}
  (\bibinfo{year}{2021}), \bibinfo{pages}{1--32}.
\newblock


\bibitem[\protect\citeauthoryear{Rankin and Irish}{Rankin and Irish}{2020}]%
        {rankin2020seat}
\bibfield{author}{\bibinfo{person}{Yolanda~A Rankin} {and}
  \bibinfo{person}{India Irish}.} \bibinfo{year}{2020}\natexlab{}.
\newblock \showarticletitle{A Seat at the Table: Black Feminist Thought as a
  Critical Framework for Inclusive Game Design}.
\newblock \bibinfo{journal}{\emph{Proceedings of the ACM on Human-Computer
  Interaction}} \bibinfo{volume}{4}, \bibinfo{number}{CSCW2}
  (\bibinfo{year}{2020}), \bibinfo{pages}{1--26}.
\newblock


\bibitem[\protect\citeauthoryear{Rankin and Thomas}{Rankin and Thomas}{2020}]%
        {rankin2020intersectional}
\bibfield{author}{\bibinfo{person}{Yolanda~A Rankin} {and}
  \bibinfo{person}{Jakita~O Thomas}.} \bibinfo{year}{2020}\natexlab{}.
\newblock \showarticletitle{The intersectional experiences of Black women in
  computing}. In \bibinfo{booktitle}{\emph{Proceedings of the 51st ACM
  Technical Symposium on Computer Science Education}}.
  \bibinfo{pages}{199--205}.
\newblock


\bibitem[\protect\citeauthoryear{Rastogi, Nagappan, and Gousios}{Rastogi
  et~al\mbox{.}}{2016}]%
        {rastogi2016geographical}
\bibfield{author}{\bibinfo{person}{Ayushi Rastogi}, \bibinfo{person}{Nachiappan
  Nagappan}, {and} \bibinfo{person}{Georgios Gousios}.}
  \bibinfo{year}{2016}\natexlab{}.
\newblock \bibinfo{booktitle}{\emph{Geographical bias in GitHub: Perceptions
  and reality}}.
\newblock \bibinfo{type}{{T}echnical {R}eport}.
\newblock


\bibitem[\protect\citeauthoryear{Rastogi, Nagappan, Gousios, and van~der
  Hoek}{Rastogi et~al\mbox{.}}{2018}]%
        {rastogi2018relationship}
\bibfield{author}{\bibinfo{person}{Ayushi Rastogi}, \bibinfo{person}{Nachiappan
  Nagappan}, \bibinfo{person}{Georgios Gousios}, {and}
  \bibinfo{person}{Andr{\'e} van~der Hoek}.} \bibinfo{year}{2018}\natexlab{}.
\newblock \showarticletitle{Relationship between geographical location and
  evaluation of developer contributions in github}. In
  \bibinfo{booktitle}{\emph{Proceedings of the 12th ACM/IEEE International
  Symposium on Empirical Software Engineering and Measurement}}.
  \bibinfo{pages}{1--8}.
\newblock


\bibitem[\protect\citeauthoryear{Robles, Arjona~Reina, Serebrenik, Vasilescu,
  and Gonz{\'a}lez-Barahona}{Robles et~al\mbox{.}}{2014}]%
        {robles2014floss}
\bibfield{author}{\bibinfo{person}{Gregorio Robles}, \bibinfo{person}{Laura
  Arjona~Reina}, \bibinfo{person}{Alexander Serebrenik},
  \bibinfo{person}{Bogdan Vasilescu}, {and} \bibinfo{person}{Jes{\'u}s~M
  Gonz{\'a}lez-Barahona}.} \bibinfo{year}{2014}\natexlab{}.
\newblock \showarticletitle{FLOSS 2013: A survey dataset about free software
  contributors: challenges for curating, sharing, and combining}. In
  \bibinfo{booktitle}{\emph{Proceedings of the 11th Working Conference on
  Mining Software Repositories}}. \bibinfo{pages}{396--399}.
\newblock


\bibitem[\protect\citeauthoryear{Robles, Reina, Gonz{\'a}lez-Barahona, and
  Dom{\'\i}nguez}{Robles et~al\mbox{.}}{2016}]%
        {robles2016women}
\bibfield{author}{\bibinfo{person}{Gregorio Robles},
  \bibinfo{person}{Laura~Arjona Reina}, \bibinfo{person}{Jes{\'u}s~M
  Gonz{\'a}lez-Barahona}, {and} \bibinfo{person}{Santiago~Due{\~n}as
  Dom{\'\i}nguez}.} \bibinfo{year}{2016}\natexlab{}.
\newblock \showarticletitle{Women in free/libre/open source software: The
  situation in the 2010s}. In \bibinfo{booktitle}{\emph{IFIP International
  Conference on Open Source Systems}}. Springer, \bibinfo{pages}{163--173}.
\newblock


\bibitem[\protect\citeauthoryear{Shah}{Shah}{2006}]%
        {shah2006motivation}
\bibfield{author}{\bibinfo{person}{Sonali~K Shah}.}
  \bibinfo{year}{2006}\natexlab{}.
\newblock \showarticletitle{Motivation, governance, and the viability of hybrid
  forms in open source software development}.
\newblock \bibinfo{journal}{\emph{Management science}} \bibinfo{volume}{52},
  \bibinfo{number}{7} (\bibinfo{year}{2006}), \bibinfo{pages}{1000--1014}.
\newblock


\bibitem[\protect\citeauthoryear{Singer, Figueira~Filho, Cleary, Treude,
  Storey, and Schneider}{Singer et~al\mbox{.}}{2013}]%
        {Singer2013CSCW}
\bibfield{author}{\bibinfo{person}{Leif Singer}, \bibinfo{person}{Fernando
  Figueira~Filho}, \bibinfo{person}{Brendan Cleary}, \bibinfo{person}{Christoph
  Treude}, \bibinfo{person}{Margaret-Anne Storey}, {and} \bibinfo{person}{Kurt
  Schneider}.} \bibinfo{year}{2013}\natexlab{}.
\newblock \showarticletitle{Mutual Assessment in the Social Programmer
  Ecosystem: An Empirical Investigation of Developer Profile Aggregators}. In
  \bibinfo{booktitle}{\emph{Proceedings of the 2013 Conference on Computer
  Supported Cooperative Work}} \emph{(\bibinfo{series}{CSCW '13})}.
  \bibinfo{publisher}{ACM}, \bibinfo{address}{New York, NY, USA},
  \bibinfo{pages}{103--116}.
\newblock


\bibitem[\protect\citeauthoryear{Singh}{Singh}{2019}]%
        {singh2019women}
\bibfield{author}{\bibinfo{person}{Vandana Singh}.}
  \bibinfo{year}{2019}\natexlab{}.
\newblock \showarticletitle{Women participation in open source software
  communities}. In \bibinfo{booktitle}{\emph{Proceedings of the 13th European
  Conference on Software Architecture-Volume 2}}. \bibinfo{pages}{94--99}.
\newblock


\bibitem[\protect\citeauthoryear{Singh and Brandon}{Singh and Brandon}{2019}]%
        {singh2019open}
\bibfield{author}{\bibinfo{person}{Vandana Singh} {and}
  \bibinfo{person}{William Brandon}.} \bibinfo{year}{2019}\natexlab{}.
\newblock \showarticletitle{Open source software community inclusion
  initiatives to support women participation}. In
  \bibinfo{booktitle}{\emph{IFIP International Conference on Open Source
  Systems}}. Springer, \bibinfo{pages}{68--79}.
\newblock


\bibitem[\protect\citeauthoryear{Steinmacher, Conte, Gerosa, and
  Redmiles}{Steinmacher et~al\mbox{.}}{2015a}]%
        {steinmacher2015social}
\bibfield{author}{\bibinfo{person}{Igor Steinmacher}, \bibinfo{person}{Tayana
  Conte}, \bibinfo{person}{Marco~Aur{\'e}lio Gerosa}, {and}
  \bibinfo{person}{David Redmiles}.} \bibinfo{year}{2015}\natexlab{a}.
\newblock \showarticletitle{Social barriers faced by newcomers placing their
  first contribution in open source software projects}. In
  \bibinfo{booktitle}{\emph{Proceedings of the 18th ACM conference on Computer
  supported cooperative work \& social computing}}.
  \bibinfo{pages}{1379--1392}.
\newblock


\bibitem[\protect\citeauthoryear{Steinmacher, Robles, Fitzgerald, and
  Wasserman}{Steinmacher et~al\mbox{.}}{2017}]%
        {Steinmacher.Teenager:2017}
\bibfield{author}{\bibinfo{person}{Igor Steinmacher}, \bibinfo{person}{Gregorio
  Robles}, \bibinfo{person}{Brian Fitzgerald}, {and} \bibinfo{person}{Anthony
  Wasserman}.} \bibinfo{year}{2017}\natexlab{}.
\newblock \showarticletitle{Free and open source software development: the end
  of the teenage years}.
\newblock \bibinfo{journal}{\emph{Journal of Internet Services and
  Applications}} \bibinfo{volume}{8}, \bibinfo{number}{1} (\bibinfo{date}{05
  Dec} \bibinfo{year}{2017}), \bibinfo{pages}{17}.
\newblock
\showISSN{1869-0238}
\urldef\tempurl%
\url{https://doi.org/10.1186/s13174-017-0069-9}
\showDOI{\tempurl}


\bibitem[\protect\citeauthoryear{Steinmacher, Silva, Gerosa, and
  Redmiles}{Steinmacher et~al\mbox{.}}{2015b}]%
        {steinmacher2015systematic}
\bibfield{author}{\bibinfo{person}{Igor Steinmacher}, \bibinfo{person}{Marco
  Aurelio~Graciotto Silva}, \bibinfo{person}{Marco~Aurelio Gerosa}, {and}
  \bibinfo{person}{David~F Redmiles}.} \bibinfo{year}{2015}\natexlab{b}.
\newblock \showarticletitle{A systematic literature review on the barriers
  faced by newcomers to open source software projects}.
\newblock \bibinfo{journal}{\emph{Information and Software Technology}}
  \bibinfo{volume}{59} (\bibinfo{year}{2015}), \bibinfo{pages}{67--85}.
\newblock


\bibitem[\protect\citeauthoryear{Storey, Zagalsky, Figueira~Filho, Singer, and
  German}{Storey et~al\mbox{.}}{2016}]%
        {storey2016social}
\bibfield{author}{\bibinfo{person}{Margaret-Anne Storey},
  \bibinfo{person}{Alexey Zagalsky}, \bibinfo{person}{Fernando Figueira~Filho},
  \bibinfo{person}{Leif Singer}, {and} \bibinfo{person}{Daniel~M German}.}
  \bibinfo{year}{2016}\natexlab{}.
\newblock \showarticletitle{How social and communication channels shape and
  challenge a participatory culture in software development}.
\newblock \bibinfo{journal}{\emph{IEEE Transactions on Software Engineering}}
  \bibinfo{volume}{43}, \bibinfo{number}{2} (\bibinfo{year}{2016}),
  \bibinfo{pages}{185--204}.
\newblock


\bibitem[\protect\citeauthoryear{Terrell, Kofink, Middleton, Rainear,
  Murphy-Hill, and Parnin}{Terrell et~al\mbox{.}}{2016}]%
        {Terell-2017}
\bibfield{author}{\bibinfo{person}{J Terrell}, \bibinfo{person}{A Kofink},
  \bibinfo{person}{J Middleton}, \bibinfo{person}{C Rainear},
  \bibinfo{person}{E Murphy-Hill}, {and} \bibinfo{person}{C Parnin}.}
  \bibinfo{year}{2016}\natexlab{}.
\newblock \showarticletitle{Gender bias in open source: Pull request acceptance
  of women versus men.(Jan 2016)}.
\newblock \bibinfo{journal}{\emph{PeerJ Computer Science}}
  (\bibinfo{year}{2016}).
\newblock


\bibitem[\protect\citeauthoryear{Terrell, Kofink, Middleton, Rainear,
  Murphy-Hill, Parnin, and Stallings}{Terrell et~al\mbox{.}}{2017}]%
        {terrell2017gender}
\bibfield{author}{\bibinfo{person}{Josh Terrell}, \bibinfo{person}{Andrew
  Kofink}, \bibinfo{person}{Justin Middleton}, \bibinfo{person}{Clarissa
  Rainear}, \bibinfo{person}{Emerson Murphy-Hill}, \bibinfo{person}{Chris
  Parnin}, {and} \bibinfo{person}{Jon Stallings}.}
  \bibinfo{year}{2017}\natexlab{}.
\newblock \showarticletitle{Gender differences and bias in open source: Pull
  request acceptance of women versus men}.
\newblock \bibinfo{journal}{\emph{PeerJ Computer Science}}  \bibinfo{volume}{3}
  (\bibinfo{year}{2017}), \bibinfo{pages}{e111}.
\newblock


\bibitem[\protect\citeauthoryear{Tinto}{Tinto}{1987}]%
        {tinto1987leaving}
\bibfield{author}{\bibinfo{person}{Vincent Tinto}.}
  \bibinfo{year}{1987}\natexlab{}.
\newblock \bibinfo{booktitle}{\emph{Leaving college: Rethinking the causes and
  cures of student attrition.}}
\newblock \bibinfo{publisher}{ERIC}.
\newblock


\bibitem[\protect\citeauthoryear{Together, Friends, de~Castro~Leal, Strohmayer,
  and Kr{\"u}ger}{Together et~al\mbox{.}}{2021}]%
        {together2021activism}
\bibfield{author}{\bibinfo{person}{Reflecting Together},
  \bibinfo{person}{Sharing Experiences Among~Critical Friends},
  \bibinfo{person}{D{\'e}bora de Castro~Leal}, \bibinfo{person}{Angelika
  Strohmayer}, {and} \bibinfo{person}{Max Kr{\"u}ger}.}
  \bibinfo{year}{2021}\natexlab{}.
\newblock \showarticletitle{On Activism and Academia}.
\newblock  (\bibinfo{year}{2021}).
\newblock


\bibitem[\protect\citeauthoryear{Tourani, Adams, and Serebrenik}{Tourani
  et~al\mbox{.}}{2017}]%
        {tourani2017}
\bibfield{author}{\bibinfo{person}{Parastou Tourani}, \bibinfo{person}{Bram
  Adams}, {and} \bibinfo{person}{Alexander Serebrenik}.}
  \bibinfo{year}{2017}\natexlab{}.
\newblock \showarticletitle{Code of conduct in open source projects}. In
  \bibinfo{booktitle}{\emph{2017 IEEE 24th international conference on software
  analysis, evolution and reengineering (SANER)}}. IEEE,
  \bibinfo{pages}{24--33}.
\newblock


\bibitem[\protect\citeauthoryear{Trinkenreich, Guizani, Wiese, Sarma, and
  Steinmacher}{Trinkenreich et~al\mbox{.}}{2020}]%
        {trinkenreich2020hidden}
\bibfield{author}{\bibinfo{person}{Bianca Trinkenreich},
  \bibinfo{person}{Mariam Guizani}, \bibinfo{person}{Igor Wiese},
  \bibinfo{person}{Anita Sarma}, {and} \bibinfo{person}{Igor Steinmacher}.}
  \bibinfo{year}{2020}\natexlab{}.
\newblock \showarticletitle{Hidden Figures: Roles and Pathways of Successful
  OSS Contributors}.
\newblock \bibinfo{journal}{\emph{Proceedings of the ACM on Human-Computer
  Interaction}} \bibinfo{volume}{4}, \bibinfo{number}{CSCW2}
  (\bibinfo{year}{2020}), \bibinfo{pages}{1--22}.
\newblock


\bibitem[\protect\citeauthoryear{Vasilescu, Filkov, and Serebrenik}{Vasilescu
  et~al\mbox{.}}{2015a}]%
        {vasilescu2015perceptions}
\bibfield{author}{\bibinfo{person}{Bogdan Vasilescu}, \bibinfo{person}{Vladimir
  Filkov}, {and} \bibinfo{person}{Alexander Serebrenik}.}
  \bibinfo{year}{2015}\natexlab{a}.
\newblock \showarticletitle{Perceptions of diversity on git hub: A user
  survey}. In \bibinfo{booktitle}{\emph{2015 IEEE/ACM 8th International
  Workshop on Cooperative and Human Aspects of Software Engineering}}. IEEE,
  \bibinfo{pages}{50--56}.
\newblock


\bibitem[\protect\citeauthoryear{Vasilescu, Posnett, Ray, van~den Brand,
  Serebrenik, Devanbu, and Filkov}{Vasilescu et~al\mbox{.}}{2015b}]%
        {vasilescu2015gender}
\bibfield{author}{\bibinfo{person}{Bogdan Vasilescu}, \bibinfo{person}{Daryl
  Posnett}, \bibinfo{person}{Baishakhi Ray}, \bibinfo{person}{Mark~GJ van~den
  Brand}, \bibinfo{person}{Alexander Serebrenik}, \bibinfo{person}{Premkumar
  Devanbu}, {and} \bibinfo{person}{Vladimir Filkov}.}
  \bibinfo{year}{2015}\natexlab{b}.
\newblock \showarticletitle{Gender and tenure diversity in GitHub teams}. In
  \bibinfo{booktitle}{\emph{Proceedings of the 33rd annual ACM conference on
  human factors in computing systems}}. \bibinfo{pages}{3789--3798}.
\newblock


\bibitem[\protect\citeauthoryear{Von~Krogh, Haefliger, Spaeth, and
  Wallin}{Von~Krogh et~al\mbox{.}}{2012}]%
        {von2012carrots}
\bibfield{author}{\bibinfo{person}{Georg Von~Krogh}, \bibinfo{person}{Stefan
  Haefliger}, \bibinfo{person}{Sebastian Spaeth}, {and}
  \bibinfo{person}{Martin~W Wallin}.} \bibinfo{year}{2012}\natexlab{}.
\newblock \showarticletitle{Carrots and rainbows: Motivation and social
  practice in open source software development}.
\newblock \bibinfo{journal}{\emph{MIS quarterly}} (\bibinfo{year}{2012}),
  \bibinfo{pages}{649--676}.
\newblock


\bibitem[\protect\citeauthoryear{Wang, Wang, and Redmiles}{Wang
  et~al\mbox{.}}{2018}]%
        {ICSE-Confidence-Competence-2018}
\bibfield{author}{\bibinfo{person}{Zhedong Wang}, \bibinfo{person}{Yi Wang},
  {and} \bibinfo{person}{David Redmiles}.} \bibinfo{year}{2018}\natexlab{}.
\newblock \showarticletitle{Competence-Confidence Gap: A Threat to Female
  Developers' Contribution on GitHub}. In \bibinfo{booktitle}{\emph{ICSE 2018
  SEIS - Software Engineering in Society}}. ICSE.
\newblock


\bibitem[\protect\citeauthoryear{Wyer~Jr and Srull}{Wyer~Jr and Srull}{2014}]%
        {wyer2014advances}
\bibfield{author}{\bibinfo{person}{Robert~S Wyer~Jr} {and}
  \bibinfo{person}{Thomas~K Srull}.} \bibinfo{year}{2014}\natexlab{}.
\newblock \bibinfo{booktitle}{\emph{Advances in social cognition, Volume I: A
  dual process model of impression formation}}.
\newblock \bibinfo{publisher}{Psychology Press}.
\newblock


\end{thebibliography}
\end{document}